\documentclass[12pt]{article}%
\usepackage{amssymb}
\usepackage{amsmath}
\usepackage{graphicx}%
\usepackage{amsfonts}%
\usepackage{natbib}
\usepackage{floatrow}
\setcitestyle{authoryear,open={(},close={)},aysep={}}
\usepackage{xr}
\usepackage{hyperref}
\usepackage{fancyhdr}
\usepackage{caption,subcaption}%
\usepackage{rotating}
\usepackage{float}

\usepackage{multirow}

\makeatother
\newfloatcommand{capbtabbox}{table}[][\FBwidth]

\newcommand{\bet}{ \mbox{\boldmath $ \eta $} }
\newcommand{\btau}{ \mbox{\boldmath $ \tau $} }

\newcommand{\bnu}{ \mbox{\boldmath $\nu$} }
\newcommand{\bmu}{ \mbox{\boldmath $\mu$} }

\newcommand{\bgamma}{ \mbox{\boldmath $\gamma$} }

\newcommand{\bdelta}{ \mbox{\boldmath $\delta$} }

\newcommand{\bpi}{ \mbox{\boldmath $\pi$} }
\newcommand{\bpsi}{ \mbox{\boldmath $\psi$} }

\newcommand{\bone}{\textbf{1}}

\newcommand{\bZ}{\textbf{Z}}

\newcommand{\bG}{\textbf{G}}

\newcommand{\bs}{\textbf{s}}

\newcommand{\bV}{\textbf{V}}

\newcommand{\bX}{\textbf{X}}

\newcommand{\bY}{\textbf{Y}}

 \renewenvironment{abstract}
 {\small
  \begin{center}
  \bfseries \abstractname\vspace{-.5em}\vspace{0pt}
  \end{center}
  \list{}{
    \setlength{\leftmargin}{.5cm}%
    \setlength{\rightmargin}{\leftmargin}%
  }%
  \item\relax}
 {\endlist}
 \usepackage{titling}
\usepackage{lineno}
\usepackage{setspace}
\title{Zero-inflated Beta distribution regression modeling \\ \small{Running head: Zero-inflated Beta regression}}
\author{Becky Tang$^{1*}$, Henry A. Frye$^{2}$, Alan E. Gelfand$^{1}$, John A Silander, Jr.$^{2}$} 
\date{}

\begin{document}

\maketitle
\noindent$^1$Department of Statistical Science, Duke University, Durham, NC 27708, USA; and $^{2}$Department of Ecology and Evolutionary Biology, University of Connecticut, Storrs, CT 06269, USA \\ $^*$ Correspondence author: becky.tang@duke.edu

\begin{abstract}
A frequent challenge encountered with ecological data is how to interpret, analyze, or model data having a high proportion of zeros. Much attention has been given to zero-inflated count data, whereas models for non-negative continuous data with an abundance of 0s are much fewer. We consider zero-inflated data on the unit interval and provide modeling to capture two types of 0s in the context of a Beta regression model. We model 0s due to missing by chance through left censoring of a latent regression, and 0s due to unsuitability using an independent Bernoulli specification. We extend the model by introducing spatial random effects. We specify models hierarchically, employing latent variables, and fit them within a Bayesian framework. Our motivating dataset consists of percent cover abundance of two plant families at a collection of sites in the Cape Floristic Region of South Africa. We find that environmental features enable learning about both types of 0s as well as positive percent cover. We also show that the spatial random effects model improves predictive performance. The proposed modeling enables ecologists to extract a better understanding of an organism’s absence due to unsuitability vs. missingness by chance, as well as abundance behavior when present. 
\end{abstract}

\clearpage 
\section{Introduction}
\label{sec:Intro}
A frequent and continuing challenge encountered with ecological data is how to interpret, analyze, or model data having a high proportion of zeros. This \emph{zero-inflation problem} \citep{martin2005zero} is found in scenarios such as: the binary presence/absence of individual organisms, counts of individuals, ranked abundances as in many vegetation plot datasets, and measures of biomass or proportion of area occupied by individuals in plot-level survey or remotely sampled data sets. A wealth of literature \citep[e.g.][]{martin2005zero,
veech2016instrinsic, blascomoreno2019what} asserts that zeros in a dataset can arise from multiple sources. Such prior studies have dichotomized zeros into “true” and “false” zeros; we avoid such value labels here, preferring stochastic interpretation.

While many statistical techniques have been proposed to address zero-inflation in discrete data, particularly for animal populations \citep{veech2016instrinsic, blascomoreno2019what}, less attention has been focused on continuous data. In particular, for count data, we remind the reader of the much-employed zero-inflated Poisson model (ZIP) \citep{lambert1992zero} which introduces an additional point mass at zero. Models such as the zero-inflated negative binomial and other power series distributions, zero-inflated N-mixture models, and a general class of count transform models have also been developed \citep{hall2000zero, ghosh2006bayesian, wenger2008estimating, veech2016instrinsic, siegfried2020count}. 

The zero-inflation challenge we focus on here is modeling the percent cover of plants within sampling plots yielding data on $[0,1)$. Ecologists often assess plant community composition and abundance through visual assessments of percent cover, which has been shown to be a robust and repeatable measure of vegetation cover \citep{vanha-majamaa_digitized_2000}. Percent cover is also often calculated by commonly used ordinal scales such as the Braun-Blanquet scale \citep{van_der_maarel_transformation_2007}. In fact, our modeling can be applied to a variety of proportional cover data found in plant and animal surveys, e.g. marine invertebrates on hard substrate or coral reef communities \citep{bell_influence_1984}, leaf cover estimates for plant phenological studies \citep{xie_predicting_2018, xie_speciesspecific_2018}, and biomass scaled as a proportion of area or organ mass \citep{jenkins_national-scale_2003}. 
 Absence, i.e., a zero, can be assumed to arise from one of three sources: unsuitability, random chance, or detection error. Unsuitability results from the biotic and abiotic factors impacting plant populations. If an area is suitable for a plant species but not found, we assume that this is due to random chance (i.e. it has not yet dispersed to that site).  
The explanation of failed detection of presence is not pursued here. In general, it requires detection/non-detection data from more than one independent visit to a sample unit. Further, it is extremely unlikely in our example system which is comprised almost exclusively of evergreen perennial species in an open shrubland.  Lastly, we do not address the notion of ``1-inflation'' here. 

Percent cover provides a challenge in that the data is continuous on $[0,1)$. Though Beta regression \citep{ferrari2004beta} is a customary choice to model such data \citep{douma_analysing_2019}, it has only been recently elaborated in detail for the context of vegetation percent cover \citep{Damgaard2019UsingTB}. However, as we clarify below, these models are not adequate for what is truly zero-inflated, continuous data. 

The key issue here is that discrete distributions explicitly place point mass at $0$ while continuous distributions do not.  So, a different mechanism is required to create mass at $0$. One method of introducing zeros for continuous data on $[0,1]$  is through the left-censoring of a latent random variable at $0$. 
For example, consider a continuous variable $W$ with mass on $(-1,1)$. Assume that the actual observation is a left-censored at $0$ realization of this variable. Then, the probability of observing a $0$ becomes $P(W \leq 0)$.  This approach is similar to the Tobit model which has been long used in economics \citep{amemiya_tobit_1984} and applied in remote sensing \citep{peterson_estimating_2005}. The Tobit model introduces a regression for latent Gaussian variables and creates point mass at $0$ through left-censoring at $0$ \citep{chib1992bayes, long1997regression}. 

A different approach to modeling zero-inflated data is a hurdle model \citep{mullahy1986specification}.  Hurdle models explain a chosen measure of abundance given presence. In our setting, they model the positive percent coverages, ignoring the observed $0$'s. Separately, the observation at each site is given a binary response according to presence or absence, modeled through a binary regression. This zero-inflation for a beta distribution, using site-specific independent Bernoulli variables, has been termed a \emph{zero-inflated Beta} model in \cite{Ospina_2012} and \cite{ospina2010inflated}. They refer to this hurdle model as BEZI (BEta Zero Inflated). 
For observations on $[0,1)$, neither of the foregoing models (BEZI or left-censored) is really a zero-inflated continuous data specification.  In both cases, a single mass at $0$ is introduced. We build upon the previous literature and propose a new model that accommodates two sources of $0$ for percent cover, analogous to the zero-inflated Poisson model for counts. . 
 
For interpretation, the left-censored regression type of $0$ will be referred to as absence by chance  \citep{blascomoreno2019what}. The Bernoulli point mass at $0$ will be referred to as  absence due to unsuitability. 
Our contribution here is to formalize a zero-inflated beta regression model for percent cover. This entails a regression within a left censored Beta specification to capture absence by chance as well as a regression to capture absence due to unsuitability.  Can we separate both sources of $0$'s?  As we discuss below, the answer is yes given the use of informative priors for identifiability. 

As the environmental covariates may not be sufficient to explain all the variability in the data, introducing spatially correlated random effects is expected to provide improved model performance.  Therefore, employing the geo-coded locations of the sampling sites, we offer a spatial random effects version of the foregoing specification.   Spatial zero-inflated models have been proposed in the literature \citep[e.g.][]{agarwal2002zero, rathbun2006spatial} and we extend this work to our proposed zero-inflated Beta regression model.  This raises the question of model comparison; is the spatial model preferred?  We address this through novel model comparison for these zero-inflated distributions.  As a case study, we consider two plant families (the Restionaceae and Crassulaceae) located across a suite of sample plots in the Cape Floristic Region (CFR) in South Africa. 

The format of the paper is as follows: Section \ref{sec:data} describes the CFR dataset which motivates our modeling.  Section \ref{sec:model} presents modeling details for the nonspatial case, offers model fitting details as well as model assessment and comparison approaches, and finishes with the spatial model.  Section \ref{sec:simulation} offers simulation investigations to serve as a proof of concept and compare various models. Section \ref{sec:cfr_application} presents the results of analyses for the CFR data. 

\section{The dataset}
\label{sec:data}

We consider percent cover for species within two plant families: Crassulaceae and Restionaceae in the Cape Floristic Region (CFR). The Crassulaceae found here are typically perennial succulents that contribute to the CFR’s diversity of succulent flora \citep{born_greater_2006}. They are arid adapted plants; 
we would expect them to inhabit drier and warmer microhabitats within the CFR, likely in sites protected by fire. The Restionaceae are evergreen reed-like plants that are an iconic component of the CFR. They are found in a wide range of habitats across the nutrient poor soils of the CFR \citep{linder_african_2001}. 

\begin{figure}[ht]
    \caption{Location of study regions Cape Point and Baviaanskloof within the Cafe Floristic Region (map adapted from \cite{turpie2003economic}).}
    \centering
    \includegraphics[scale = 0.5]{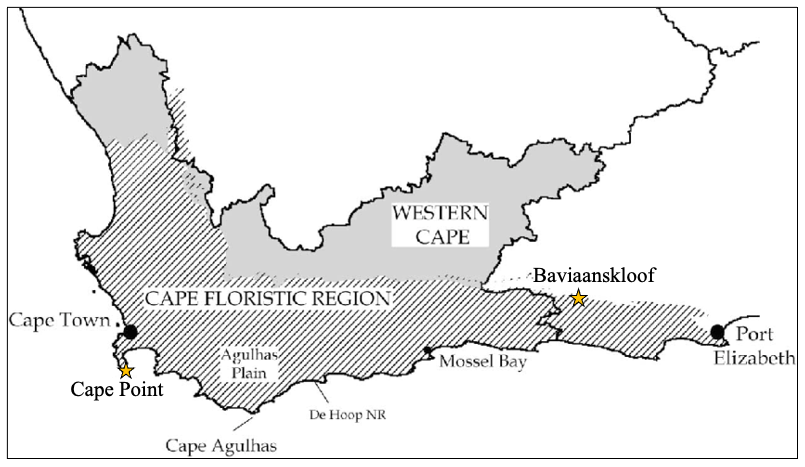}
    \label{fig:CFR_map}
\end{figure}

\begin{figure}[ht]
    \caption{Exploratory analysis of observed percent cover for Crassulaceae (pink) and Restionaceae (navy). Where appropriate, orange corresponds to observed 0 percent cover.}
    \centering
     \begin{subfigure}[b]{0.35\textwidth}
            \caption{Histograms of percent cover by species across the two regions.}
     \centering
        \includegraphics[scale = 0.55]{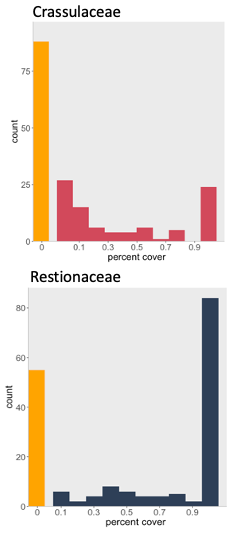} 
        \label{fig:pct_cover_hist}
     \end{subfigure} \quad
     \begin{subfigure}[b]{0.6\textwidth}
              \caption{Top: positive percent cover plotted against environmental covariates. Bottom: distribution of  environmental covariates at locations with observed 0.}
         \centering
         \includegraphics[scale = 0.45]{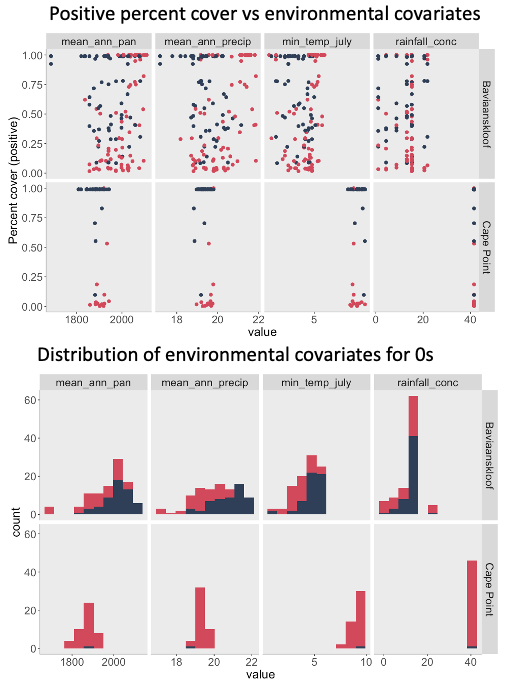}
         \label{fig:eda_covars_all}
     \end{subfigure} \\
     \begin{subfigure}{1\textwidth}
     \centering
            \caption{Spatial maps of percent cover within Baviaanskloof.}
        \includegraphics[scale = 0.5]{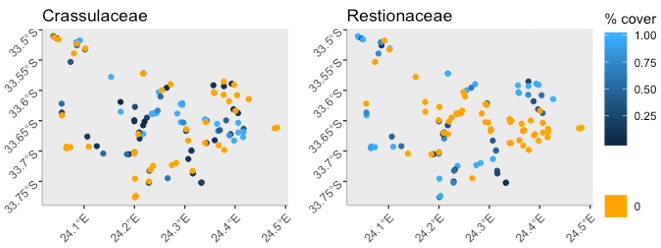}
   \label{fig:spatial_bav_pcts}
     \end{subfigure}
    \label{fig:eda}
\end{figure}

Our data were collected between 2010 and 2011 from plots concentrated in two regions of South Africa: 61 plots in the Cape Point section of Table Mountain National Park and 119 plots in the Baviaanskloof Mega-Reserve (Fig. \ref{fig:CFR_map}). Following the relevé method \citep{ellenberg_community_1974}, a well-established inventory protocol used to classify vegetation and inventory species 
\citep[e.g][]{schaminee2009vegetation, dengler2011global}, each 10 x 5 m  plot was sampled for its floristic composition using observations of percent cover. 
Histograms of the observed percents are presented in Fig. \ref{fig:pct_cover_hist}. 49\% of the observations for Crassulaceae and 31\% of observations for Restionaceae are $0$, arguing for our proposed introduction of zero-inflation. 

The plots are associated with a variety of environmental variables georeferenced for each plot and largely derived from the WorldClim \citep{hijmans_very_2005} and Schulze \citep{schulze_south_1997} climate datasets. We employ four of these variables in our regression modeling: mean annual evapotranspiration (1-minute grid cell), mean annual precipitation (30 arc second), average minimum temperature for July (the austral winter; 1-minute grid cell), rainfall concentration index (1-minute grid cell; scaled 0 (rainfall is evenly divided across all months) to 100 (rain only falls one month of the year)).  Fig. S\ref{fig:eda_covars} displays histograms of the environmental covariates, and Fig. \ref{fig:eda_covars_all} plots the positive percent covers of both species against the environmental covariates as well as the distribution of environmental covariates for observed 0s. As our observations are spatially referenced, we may also be interested in investigating if the data are spatially correlated for a given family.
The percent covers in Baviaanskloof and Cape Point plotted by longitude-latitude are presented in Figs \ref{fig:spatial_bav_pcts} and S\ref{fig:spatial_cp_pcts}, respectively. Throughout this work, `S' denotes Supplementary Information.

\section{Building the model}
\label{sec:model}
As we have discussed above, we propose a site-level model that will have a Beta distribution specification for the continuous observations. It will employ left censoring to create a point mass at $0$ which corresponds to missingness by chance at the site.  It will introduce a Bernoulli trial at the site to capture the probability of unsuitability.

\subsection{Left-censored Beta regression}
\label{subsec:left-censor}
\begin{figure}[ht]
    \caption{Example of the LECB model. A Beta(3,2) distribution is extended to (-1,1) using $a= 1$ and subsequently left-censored. \textbf{a)} Probability and cumulative distribution functions. \textbf{b)} Histograms of randomly generated data.}
    \centering
    \begin{subfigure}{0.95\textwidth}
        \caption{PDF of Beta(3,2) (left) and PDF (middle) and CDF (right) of Extended Beta(3,2). Dark-grey shaded area denotes the left-censoring associated with the Beta mass at 0.}
     \includegraphics[scale = 0.5]{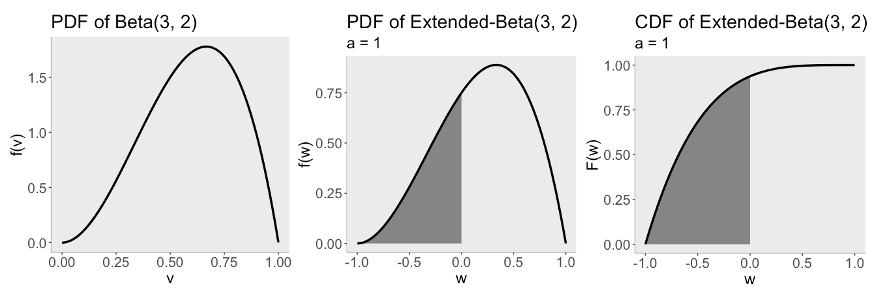}
     \label{fig:extended_pdf_cdf}
    \end{subfigure}\\
        \begin{subfigure}{0.9\textwidth}
        \centering
    \caption{Histograms of data generated under the Beta(3,2) distribution (left) and subsequently extended and left-censored (right).}
     \includegraphics[scale = 0.5]{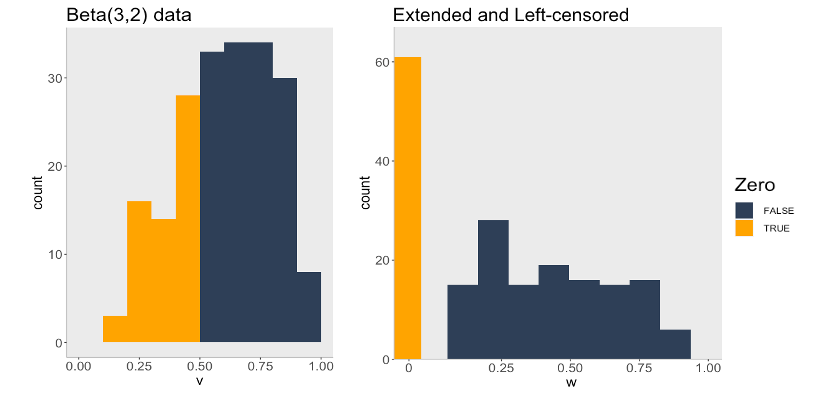}
     \label{fig:extended_beta_data}
    \end{subfigure}
    \label{fig:extended_beta}
\end{figure}

We first introduce a $0$ through left censoring. In a general setting, for site $i$, let the observation, $Y_i = \text{max}(0,W_i)$ where $W_i$ is a continuous variable on $(-\infty, \infty)$ with distribution $f_i$ so
$P(Y_i=0) = P(W_i \leq 0) = \int_{-\infty}^{0} f_{i}(w)dw$.  When $f_i$ is a normal distribution this model is a Tobit model \citep{tobin1958estimation}. In our context, we take $f_i$ to be the Beta distribution.  However, the Beta has bounded support on $(0,1)$. Therefore, we propose extending the Beta distribution: if $V \sim \text{Beta}(\alpha, \beta)$, then for $0< a < \infty$, $W = (a+1)V -a$ provides an extended or re-scaled Beta on $(-a, 1)$. The probability of a negative $W$ provides our Beta mass at 0: $P(W \leq 0) = P(V \leq \frac{a}{a+1} \equiv c)$ (Fig. \ref{fig:extended_beta}).

We follow a Bayesian approach for modeling fitting. We introduce a latent Beta variable $V_i$ such that $Y_i = \max\{0, W_i = (a+1)V_i - a\}$. Then for all $Y_i > 0$, we immediately know $V_i$. For $Y_i =0$, the associated $V_i$ is less than or equal to $c$ by construction. Thus, introducing the latent $V_i$ allows us to have the `complete' data. In fitting the model, we work with a convenient reparametrization for Beta regression \citep{ferrari2004beta}. We employ $\mu_i = \alpha_{i}/(\alpha_{i} + \beta_{i})$, the mean parameter and $\nu_i = \alpha_i + \beta_i$, the so-called ``sample size'' parameter. The distribution of the extended Beta on $(-a,1)$ is also parameterized in terms of $\mu_i$ and $\nu_i$.  We model $logit(\mu_i) =  \bX_i^T \bdelta$, and  $log(\nu_i) = \bX_i^T \bpsi$ where $\bX_{i}$ is a vector of environmental covariates associated with site $i$.

With regard to addressing the effect of choice of $a$, the Appendix offers some analytical investigation. The choice of $a=1$ mirrors the Tobit model by providing matching support above and below 0. Ideally, we would take $a$ to be a model parameter and learn about its value during model fitting, along with $\bdelta$ and $\bnu$. In implementation, this leads to identifiability issues:  $a$ (or equivalently, $c$) and the intercept  $\delta_{0}$ for $\mu$ compete to explain the probability of 0. One approach to address this issue would be to use an informative prior on $c$ or $\delta_{0}$. However, we remark that the value of $c$ does not have interpretation; it is a mechanism used to obtain $0$'s. Practically, this suggests holding $c$, equivalently $a$, fixed at a pre-specified value during model fitting and then doing out-of-sample model comparison. 

In this regard, and following the Appendix, through simulations we briefly explore sensitivity of inference to the choice of $a$. The simulations are performed by generating data under a $c_{\text{true}}$, fitting the censored model with several different  values of $c_{fixed}$, and examining how recovery of the covariate effects and probabilities of 0 are affected. The regression for $\mu$ is $logit(\mu) = \delta_{0} - 0.5x_{1}$,  with different $\delta_{0}$ and $\psi$ in each simulation. We evaluate credible interval coverage of $\delta_{1}$ and root mean square predictive error (RMSPE) of the probabilities of 0 across 100 simulations for each pair $(c_{\text{true}}, c_{\text{fixed}})$.  We find that generally, recovery of the probability of 0 is robust to the choice of fixed $a$; predicted probabilities on the held-out test set tend to yield similar RMSPEs (Fig. S\ref{fig:rmspe_c}). The largest differences in predictions and RMSPEs occur when $c_{\text{true}}$ is small and $c_{\text{fixed}}$ is large, or vice versa (Fig. S\ref{fig:censor_sim1_prob0}).

Empirical coverage of the credible intervals for $\delta_{1}$ achieves or is slightly below nominal when $c$ is fixed at the true value, but tends to decrease as the difference between $c_{\text{true}}$ and $c_{\text{fixed}}$ increases (Fig. \ref{fig:emp_cvg_c}). This decreasing coverage is exacerbated when data are generated with smaller intercept $\delta_{0}$, holding everything else fixed (ex. Simulations 1-3, Simulations 4-6). This is due to the larger number of 0s in the data induced by the smaller intercept.
However, in almost all pairs of $(c_{\text{true}}, c_{\text{fixed}})$, the sign of $\delta_{1}$ is correctly estimated (Fig. S\ref{fig:prop_sign_c}). 
Therefore, even if $c$ is fixed at an incorrect value, the model is expected to recover the correct covariate relationships.

From the simulations, unsurprisingly, we find that fixing $c$ at the truth leads to the best recovery of parameters and predicted probabilities of 0. We generally find that fixing $c$ to be large when the data contain a high proportion of 0s leads to better performance, and similarly for small $c$ with a smaller proportion of 0s. Inference is the least robust when the magnitude of the difference between $c_{\text{true}}$ and $c_{\text{fixed}}$ is large. Thus, if the modeler seeks to choose a single value $\textit{a priori}$, perhaps fixing $c = 0.5$ ($a=1$) is the most natural option. Again, we suggest analyzing the data with differences choices of $c$ and performing sensitivity analysis to determine the best fitting model. For the remainder of the text, we refer to this left-censored extended Beta with single-zero model as `LCEB'.

\subsection{Zero-inflated Beta modeling - the nonspatial case}
\label{sec:mod_nonspatial}
Let $Z_i$ be a Bernoulli variable with $P(Z_i = 1) = \pi_i$ and let $f_i$ be a density on $(0,1)$.  Then, suppose $Y_i =0$ if $Z_i =1$, and $Y_i \sim f_i(y)$ if $Z_i=0$.  Here, $P(Y_i = 0) = \pi_i$. This is the BEZI model of \cite{ospina2010inflated}. It is a hurdle model analogous to the setting where $Y_i$ is a count variable and $f_i$ is the Poisson distribution truncated at 1. \citep{mullahy1986specification}.  

If $Z_i=0$, then let $Y_i$ be a realization of the censored Beta regression model described in Section \ref{subsec:left-censor}: $Y_i = \text{max}(0,W_i)$ where $W_i = (a+1)V_{i} - a$ and, using the \cite{ferrari2004beta} parameterization, $V_{i} \sim Beta(\mu_{i}, \nu_{i})$.  Now, $P(Y_i=0) = P(Z_i=1) + P(Z_{i}=0 \cap W_{i} \leq 0) = \pi_i + (1-\pi_{i})\int_{-a}^{0}f_{i}(w)dw$ (so $P(Y_i >0)= (1-\pi)\int_{0}^{1}f_{i}(w)dw$).  We immediately see the two sources of $0$'s as well as the parallel construction with the familiar zero-inflated count model.  In this regard, we will refer to $(1-\pi_{i})\int_{-1}^{0}f_{i}(w)dw$ as the probability of absence by chance with $\pi_i$ the inflated probability of absence which we ascribe to unsuitability.

We can identify the two sources of $0$'s if $a$ is held fixed because the $\mu_i$ and $\nu_i$ are informed by all of the data while only the $0$'s inform about $\pi_i$.  However, informative prior information is needed to capture the relative magnitudes of the chances of each source. In the absence of further knowledge, this is addressed via out-of-sample model comparison. We refer to our proposed zero-inflated Beta model as $\mathcal{M}_{1}$.

Again, following the above, for all $Y_i > 0$, we immediately know $V_i$. For $Y_i =0$, we know that if the observation arose from the Beta, by construction, the associated $V_i$ is less than or equal to $\frac{a}{1+a} = c$. Our zero-inflated model takes the form
\begin{equation}
  \begin{array}{l}
 P(Y_i = 0 | \pi_i, \mu_i, \nu_i) = \pi_i + (1-\pi_i) P(V_i \leq c|  \mu_i, \nu_i) \\       f(y_i | \pi_i,  \mu_i, \nu_i) = (1-\pi_i) f_{v_i}(v_i | \mu_i, \nu_i), \quad y_i > 0
  \end{array}
\end{equation}

Here, a latent indicator variable $Z_i$ is introduced to determine the source of an observation: if $Z_i = 1$ then the associated $Y_i$ arises as a 0 from the degenerate process, and if $Z_i = 0$ then $Y_i$ is a realization of the extended Beta. Then we have that $P(Y_i = 0, Z_i = 1 | \pi_i,  \mu_i, \nu_i) = \pi_i$ and $P(Y_i = 0, Z_i = 0 | \pi_i,  \mu_i, \nu_i) = (1-\pi_i) P(V_i \leq c |  \mu_i, \nu_i)$. 

Given the parameters $\pi_i, \mu_i, \nu_i$ and the latent $Z_i$, the likelihood is:
\begin{equation}
\begin{array}{ll}
      L(\bpi ,\bmu, \bnu, \bZ; \bY ) &= 
      \prod_{i=1}^{n} \pi_i^{z_i} \cdot \left((1-\pi_i) f_{y_i}(y_i |  \mu_i, \nu_i) \right)^{1-z_i}.
\end{array}
\end{equation}

We model $\text{logit} \pi_i = \bG_i^T \bgamma$, and as in the case of the single-zero censored Beta, $\text{logit}\mu_i =  \bX_i^T \bdelta$, and  $\log \nu_i = \bX_i^T \bpsi$ where $\bX_{i}, \bG_{i}$. The context will determine the commonalities between $\bX_{i}$ and  $\bG_i$ with regard to explaining $\pi$ and the parameters of the Beta distribution. Then, we wish to infer about the regression coefficients, $\bgamma, \bdelta, \bnu$, as well as the latent $Z_i$ associated with $Y_i=0$. With priors, the posterior distribution is proportional to
\begin{equation}\label{eq:posterior}
    \prod_{i=1}^n f_{Y_i | Z_i, \bdelta, \bnu} (Y_i | Z_i, \bdelta,\bnu ) \cdot  \prod_{i=1}^n f_{Z_i | \bgamma}(Z_i | \bgamma) \cdot f_{\bgamma}(\bgamma) \cdot f_{\bdelta}(\bdelta) \cdot f_{\bpsi}(\bpsi).
\end{equation} 
Prior choices follow in the next subsection.

\subsection{Model fitting}
\label{sec:mod_fitting}

We use a Gibbs sampler with Metropolis updates for the $\bdelta, \bgamma, \bpsi$. If $Y_i > 0$, we immediately have $Z_i$ and $V_i$. Then for all $i$ where $Y_i = 0$, at every iteration within the sampler we sample $Z_i$ from its Bernoulli full conditional with success probability $P(Z_i = 1| Y_i = 0) =\pi_i/(\pi_i  + (1-\pi_i) P(V_i \leq c))$.  If $Z_i = 0$, we sample an associated latent $V_i$ from its conditional Beta distribution restricted to the interval $(0,c)$, analogous to sampling from truncated Normal variables in Bayesian Tobit modeling \citep{chib1992bayes}. Given  $\bZ$ and $\bV$, the parameters $\bgamma$ are independent of the parameters $(\bdelta, \bpsi)$.  In the sequel, we assume $\nu_i = \nu$, i.e., a common sample size parameter across sites.  This is in accord with \cite{Damgaard2019UsingTB} who use the parameterization, $\mu_i$ with common shape parameter $\delta= 1/(\nu +1)$.  

Distinguishing the two types of zeros in the data is challenging. The magnitudes of $\pi_i$ and $\mu_i$ are strongly influenced by their intercepts $\gamma_0$ and $\delta_0$, respectively. A large positive $\gamma_0$ will result in $\pi_i$ close to one, encouraging unsuitability absence; a large negative $\delta_{0}$ will result in $\mu_i$ small, encouraging absence by chance. Recall that the probability of a zero at location $i$ is $\pi_i + (1-\pi_i) P(V_i \leq c | \mu_i, \nu_i)$. The second term leads to difficulty in separating the source of a zero due to weak identifiablity of the intercepts. Rather than removing an intercept from $\pi_i$ or $\mu_i$, we choose to adopt a strong prior on one of the intercepts.

This leads to two questions: which intercept, and what prior? The natural choice is the intercept $\gamma_0$ for the incidence of inflated zeros; an ecologist may have a general belief of how (un)favorable a location is for a given species, but may have less knowledge about the expected percent cover. If so, a tight prior centered at the modeler's belief for $\gamma_0$ can be used. We take this approach in our simulations. However, we suggest performing modeling comparison or sensitivity analysis with different priors to select a model. Our proposed metrics for modeling comparison follow in the next subsection. We use non-informative or weakly-informative priors for all the remaining parameters. For a full list of priors used in all models, please see Table S\ref{tab:prior_choices} in the Supplement.

\subsection{Model assessment and comparison approaches}
\label{sec:mod_assess}

If the source of a zero is known--as with the simulated data below--one measure of model assessment is through classification of the type of zero. We use receiver operating characteristic (ROC) curves and the corresponding area under the curve (AUC) for the true source of a zero versus the predicted probability of arising from that source. In practice, we won't know the true source of a zero. Instead, we assess and compare models in terms of their ability to distinguish zero and positive observations. Another metric is Tjur's $R^2$ statistic, a coefficient of discrimination often used to evaluate logistic regression prediction \citep{tjur2009coefficients}. This statistic is calculated as $R^2 = \bar{\hat{p}}_1 - \bar{\hat{p}}_0$, where $\bar{\hat{p}}_1$ and $\bar{\hat{p}}_0$ are the average of fitted values for successes and failures. 
Larger AUC and $R^2$ indicate superior classification performance.

The continuous ranked probability score (CRPS) which compares the entire predictive cumulative distribution function, $F(x)$, to an observed held-out data point \citep{gneiting2007strictly} offers further model comparison. For a continuous CDF, $CRPS = \int (F(x) - \bone( y  < x ))^2 dx$, where $y$ is the observed value. For a discrete distribution, the ranked probability score replaces the integral with a sum. A smaller (C)RPS indicates that the predictive distribution is more concentrated around the observation and is thus preferred.  In the case of our proposed distribution, we have a continuous predictive CDF with a point mass at zero. Let $p_0$ denote the mass at 0. 
Then our score takes the form:
$$ (C)RPS = \begin{cases}
    p_0^2 + \int_{0}^{\infty} (F(x) - \bone( y  < x ))^2 dx, \qquad y > 0\\
    (1-p_0)^2 + \int_{-\infty}^{0} (1-F(x))^2 dx, \qquad y = 0
    \end{cases}
$$

Thus, we can obtain the (C)RPS for the full distribution, denoted as $CRPS_f$. How well the predictive distribution approximates the positive observations may also be of interest. In this case, we omit the zero observations from the calculation and only use the positive data. We refer to this resulting score as $CRPS_h$. We use Monte Carlo integration to evaluate the integrals (Supplement S\ref{sec:mc_integration}). 

Because evaluation using our ranked probability score requires separating observations based on whether they are zero or positive, the concentration of the predictive distribution depends on the number of sources of zeros that are being modeled. For example, it would be inappropriate to use our $CRPS_f$ to compare our proposed model with the BEZI model. However, if we wish to perform sensitivity analysis to assess the effect of prior choice in our proposed model, then the use of $CRPS_f$ would be appropriate to compare models. 
\subsection{Zero-inflated Beta modeling - the spatial case}
\label{subsec:mod_spatial}

With observations recorded at geo-coded (point-referenced) locations, spatial dependence can be introduced using random effects.  Within our zero-inflated Beta regression model, we can insert these spatial random effects in two places. The first is in the mean specification, e.g., $\text{logit} \mu (\bs_i) = \bX_i^T \bdelta + \eta(\bs_i)$, where $\eta(\bs_i)$ is the random effect at the geo-coded location $\bs_i$. In this case, the random effect impacts both the positive percent covers as well as the classification of the zero source. The second introduces a spatial random effect $\tau(\bs_i)$ into the probability that a site is unfavorable, e.g., $\text{logit} \pi(\bs_i) = \bG_i^T \bgamma + \tau(\bs_i)$. Then, space impacts the classification of the zero source.
Given an unobserved set of locations and associated predictors, we can perform spatial prediction for percent covers at these locations.

We model the set of spatial random effects $\eta(\bs)$ using a Gaussian process centered at $0$ with exponential covariance function, $C(\bs_{i}, \bs_{i} + h) = \sigma^2 e^{ -\phi_\eta || h||}$. As shown by \cite{zhang2004inconsistent}, the product of the spatial variance $\sigma^2$ and decay term $\phi$ is identifiable, but the individual terms are not. With greater interest in learning about spatial variability , we set $\phi$ to be fixed during model fitting and estimate $\sigma^{2}$  (see \cite{banerjee2014hierarchical} in this regard). The posteriors for both spatial models are provided in Supplement S\ref{sec:spatial_posts}.

We fit the $\eta(\bs)$ ($\tau(\bs)$) with an  elliptical slice sampler \citep{murray2010elliptical}. As mentioned previously, we hold the spatial decay parameter fixed and estimate the spatial variance $\sigma^{2}$ for identifiability \citep{zhang2004inconsistent}. We take a weakly informative prior Inverse Gamma(0.5, 0.5) prior on $\sigma^{2}_{\eta}$ ($\sigma^{2}_{\tau}$). Introducing a random effect in the regression to create $\pi(\bs_i)$ or $\mu(\bs_i)$ may lead to instability when estimating the corresponding intercept. Therefore, we suggest using an informative prior for the intercept in the regression where the random effect is introduced, and performing model assessment and comparison to determine the best model. For the remainder of the paper, let $\mathcal{M}_2$ denote the model where the spatial random effect $\eta(\bs_i)$ is inserted into the mean for the extended Beta, and  $\mathcal{M}_3$ the model where the random effect $\tau(\bs_i)$ is included in the probability of unsuitability. Including the spatial process in models $\mathcal{M}_{2}$ and $\mathcal{M}_{3}$ leads to increased computational time compared to the non-spatial $\mathcal{M}_{1}$, especially as the size of the data set increases (ex. about 0.5 vs. 3 minutes for $n = 300$ using a server node with one core  Intel(R) Xeon(R) CPU E5-2680 v3 @ 2.50GHz; Fig. S\ref{fig:run_times}).

\section{Simulation}
\label{sec:simulation}

\subsection{$\mathcal{M}_{1}$: a proof of concept}
\label{sec:nonspatial_sim}

We generate data under our proposed zero-inflated Beta model $\mathcal{M}_{1}$, and examine how well we recover the truth. Data are generated and models fit with $a=1$. We include a $N(0,1)$ covariate for $\pi_i$, and a separate $N(0,1)$ variable for $\mu_i$ and have a constant $\nu_i = \nu$. For the intercept in unsuitability, we use an informative Normal prior centered at the true $\gamma_0$ with  0.25 standard deviation. We run the samplers for 7500 iterations after 2500 burn-in, and thin to retain every fifth sample. We examine traceplots and autocorrelation to assess convergence. All analyses were conducted using R (Version 3.6.1) \citep{team2013r}. 

As noted above, the intercepts in $\pi_i$ and $\mu_i$ largely control the amount of zeros in the data, as well as the proportions of the sources of zeros: Simulation 1.1 has equal proportions, Simulation 1.2 has more zeros arising from the Beta, and Simulation 1.3 has more zeros arising due to unsuitability. 
Plots of ROC curves demonstrate that our model is able to perform much better than random guessing in separating the two types of zeros (Fig. S\ref{fig:m1_roc}).


\subsection{Model comparison: nonspatial models}
\label{sec:nonspatial_comp}

We compare BEZI, LCEB, and $\mathcal{M}_{1}$ by generating train and test data under each of these models, fitting all three models on the sets of train data, and predicting on the sets of test data. 
In the case of  LCEB, and $\mathcal{M}_{1}$, $a$ is fixed at 1.
We use Tjur's $R^2$ and AUC to evaluate the ability of each model to distinguish zero and positive observations, and $CRPS_h$ to compare each model's predictive distributions for the positive observations. Comparisons are performed for varying amounts of 0s in the data and, for data generated under $\mathcal{M}_{1}$, with different proportions for the sources of 0s by changing $\delta_{0}$ and $\gamma_{0}$. 
We perform 50 replications for each data-generating model and set of $\delta_{0}$ and $\gamma_{0}$, and present results averaged across the 50 runs.  More details on the simulations are provided in the Supplement S\ref{sec:simulation_details}.

When the data-generating model is $\mathcal{M}_{1}$, the true model outperforms the other models using all three prediction metrics (Table \ref{tab:m1_full_sim}). The higher $R^2$ and AUC indicate its superior ability in separating the positive observations from the zero observations. As evidenced by the lower $CRPS_h$,  $\mathcal{M}_1$'s predictive distribution for positive observations is more concentrated around the truth in all scenarios. 
These findings reveal that the BEZI and LCEB models cannot capture all the $0$'s well enough when data are generated with two sources. That is, the regression for $\pi$ in BEZI cannot explain all the $0$'s, and the regressions for $\mu$ and $\nu$ in LCEB cannot accommodate both the positive percent covers and the large incidence of $0$'s.

When BEZI is the data-generating model, we see that model $\mathcal{M}_{1}$ performs very similar to and sometimes better than the true model (Table S\ref{tab:bezi_full_sim}). In particular, when $\delta_{0}$ is large and positive (i.e. the mean of the positive percent covers is large), the model $\mathcal{M}_{1}$ sometimes outperforms the true model. However, when $\delta_{0}$ is negative, the BEZI model offers prediction performance. This can be attributed to the dual role that $\delta_{0}$ plays in $\mathcal{M}_{1}$, as $\delta_{0}$ helps determine the mean of the positive percent covers and also contributes to the probability of 0. Under the BEZI model, $\delta_{0}$ only has a single role. Thus when $\delta_{0}$ is negative, the probability of 0 must increase under $\mathcal{M}_{1}$ if the model wishes to capture the lower positive percent covers.

If the left-censoring model LCEB generates the data, we find that $\mathcal{M}_{1}$  outperforms or performs just as well as the true model in almost all scenarios (Table S\ref{tab:cens_full_sim}). Additionally, the BEZI model often performs just as well.
This is explained by the extra $\pi$ parameter in both the $\mathcal{M}_{1}$   and BEZI models, as it provides flexibility to accommodate the behavior of 0s generated under the censor-only model. That is,  the likelihood under LCEB is $L(a, \bdelta, \bpsi; \{Y_{i}\})$, while under BEZI it is  $L(\bgamma, \bdelta, \bpsi; \{Y_{i}\})$ and under $\mathcal{M}_{1}$ it is $L(a, \bgamma, \bdelta, \bpsi; \{Y_{i}\})$. For this reason, LCEB has poor performance for data generated under BEZI and $\mathcal{M}_{1}$, precisely because it has fewer parameters and thus less flexibility (Tables \ref{tab:m1_full_sim}, S\ref{tab:bezi_full_sim}).

\begin{table}[ht]
    \caption{Simulation for data generated under the proposed zero-inflated Beta model $\mathcal{M}_{1}$. $R^2$ and AUC for classifying between zero and positive, and $CRPS_h$ for predictive distribution of positive observations for various simulations. Simulations vary by amount of zeros in the data, as well as the proportions of zeros arising from each source (\% unsuitable, \% random chance). Bold indicates best performer.}
    \centering
    \resizebox{\textwidth}{!}{%
    \begin{tabular}{|c|ccc|ccc|ccc|}
    \hline
    &\multicolumn{3}{c|}{$R^2$}&\multicolumn{3}{c|}{AUC} &\multicolumn{3}{c|}{$CRPS_h$} \\ \hline
   \small{(\% deg. 0, \% Beta 0)} & BEZI & LCEB & $\mathcal{M}_{1}$  & BEZI & LCEB & $\mathcal{M}_{1}$ & BEZI & LCEB & $\mathcal{M}_{1}$  \\ \hline
    18\%, 5\% & 0.046 & 0.033 & \textbf{0.064} & 0.622 & 0.590 & \textbf{0.656} & 0.085 & 0.115 & \textbf{0.058}\\ 
    12\%, 12\% & 0.016 & 0.092 & \textbf{0.096} & 0.574 & 0.647 & \textbf{0.667} & 0.138 & 0.083 & \textbf{0.073} \\ 
    5\%, 19\% & 0.002 & 0.160 & \textbf{0.168} &  0.561 & 0.746 & \textbf{0.750} & 0.144 & 0.080 & \textbf{0.077} \\ 
    \hline \hline
    38\%, 15\% & 0.074 & 0.034 & \textbf{0.097} & 0.634 & 0.604 & \textbf{0.672} & 0.125 & 0.150 & \textbf{0.083}\\ 
    22\%, 26\% & 0.007 & 0.152 & \textbf{0.158} & 0.563 & 0.716 & \textbf{0.719} & 0.148 & 0.097 & \textbf{0.084}\\ 
    13\%, 37\%& 0.003 & 0.193 & \textbf{0.198} & 0.560 & 0.752 & \textbf{0.753} & 0.145 & 0.096 & \textbf{0.087} \\
          \hline \hline
    71\%, 9\% & 0.046 & 0.025 & \textbf{0.051} & 0.600 & 0.594 & \textbf{0.647} & 0.151 & 0.191 & \textbf{0.086} \\ 
    39\%, 40\% & 0.014 & 0.125 & \textbf{0.135} & 0.579 & 0.729 & \textbf{0.748} & 0.140 & 0.124 & \textbf{0.099} \\ 
    15\%, 56\% & 0.002 & 0.183 & \textbf{0.187} & 0.550 & 0.772 & \textbf{0.771} & 0.135 & 0.104 & \textbf{0.094} \\
    \hline     
    \end{tabular}
    }
    \label{tab:m1_full_sim}
\end{table}

\subsection{Model comparison: spatial vs nonspatial zero-inflated models} 
\label{sec:spatial_sim}

Again, $\mathcal{M}_{1}$ is the non-spatial zero-inflated model we developed in Section 3.2, and $\mathcal{M}_{2}$ and $\mathcal{M}_{3}$ are the spatial versions with space in the either the mean or probability of site unsuitability, respectively. To compare the spatial and nonspatial versions, we conduct simulations generating $n_{train} = 400$ observations from the spatial model, fitting the data to the true model and $\mathcal{M}_1$, and predicting on $n_{test} = 200$ points. The spatial region is a square on $(0,50) \times (0,50)$, and locations are generated randomly uniformly. Data are generated and the models fit using $a = 1$, $\sigma^{2} = 1$, and $\phi = 20$, with single independent $N(0,1)$ covariates in $\pi(\bs_{i})$ and $\mu(\bs_{i})$.   We use a Normal prior centered at the truth with standard deviation 0.25 for the appropriate intercept when fitting either the true model ($\mathcal{M}_2$ or $\mathcal{M}_3$) and $\mathcal{M}_1$. When fitting $\mathcal{M}_{2}$ or $\mathcal{M}_{3}$, $\phi$ is held fixed at one-third the maximum observed distance.

We first compare $\mathcal{M}_{2}$ with $\mathcal{M}_{1}$. Simulations 2.1-2.3 in the Supplemental Information compare the parameter recovery of the two models for $\sigma^{2}_{\eta}= \{0.5, 1, 2\}$. Table  S\ref{tab:re1_post_sum} of posterior means and 95\% credible intervals reveal the ability of $\mathcal{M}_{2}$ to recover the true parameters, including the spatial variance.  Fig. S\ref{fig:sim2_re} shows that we are able to estimate the spatial surface well. In contrast, we find $\mathcal{M}_{1}$ often fails to capture the truth for the parameters involved in the Beta regression as $\sigma^{2}_{\eta}$ increases. 
We then evaluate the predictive performance of the two models by generating data under $\mathcal{M}_{2}$ with varying proportions of 0s, and evaluating $R^2$, AUC, and $CRPS$ for when fitting the data with $\mathcal{M}_{2}$ and $\mathcal{M}_{1}$, averaged across 50 runs (Table S\ref{tab:sims_m2}). $\mathcal{M}_2$ always outperforms in terms of AUC, $R^2$, and $CRPS_h$. Interestingly, when the data consist of a large proportion of zeros ($>70\%$, bottom right cells in Table S\ref{tab:sims_m2}), the incorrect model actually performs better in terms of $CRPS_f$. In these scenarios, the prevalence of $Y_i >0$ is sparse which inhibits learning the spatial surface.
For a $Y_i = 0$, whether $\mu(\bs_i)$ receives a random effect depends on the iteration's Bernoulli trial for the corresponding $Z_i$.
We suspect this uncertainty leads to some model instability, and reflects the weakness of $\mathcal{M}_2$ in separating the two sources with a large number of zeros in the data.

We also generate data where the spatial effect is added to the regression for $\pi(\bs_i)$ and compare $\mathcal{M}_{3}$ with $\mathcal{M}_{1}$ (Simulations 3.1-3.3, Supplement). We find that $\mathcal{M}_{3}$ and $\mathcal{M}_{1}$ perform very similarly in terms of parameter recovery (Table S\ref{tab:re2_post_sum}, S\ref{fig:re2_comp_plots}, third column). Fig. S\ref{fig:sim3_re} displays the estimated spatial surface, and S\ref{fig:re2_comp_plots} presents plots of comparisons from these simulations. Results of simulations comparing predictive performance under differing amounts of zeros are presented in Table S\ref{tab:sims_m3}. We find that $\mathcal{M}_{3}$ is marginally superior to the non-spatial model in all scenarios across the four metrics, except when there are a few number of zeros in the data (bottom right cell of Table S\ref{tab:sims_m3}). In these cases, only a fraction of the already low-prevalence $0$s will arise due to unsuitability. Therefore, the estimate of $\pi_{i}$ parameter in the non-spatial model is only marginally impacted by the lack of spatial process, which serves to inflate/deflate the probability of zero.

\section{Application to Cape Floristic Region data}
\label{sec:cfr_application}
\subsection{ $\mathcal{M}_{1}$: Results for the CFR data}
\label{sec:nonspatial_analysis}

We fit our proposed model to the percent cover data for two plant families, Crassulaceae and Restionaceae. We use average annual potential evaporation and the rainfall concentration index as predictors for unsuitability. These predictors capture the overall aridity of a site which should provide a hard threshold (particularly for the succulent Crassulaceae) as to whether certain species within the two families could establish. We use average annual precipitation and minimum temperature in July (austral winter) to estimate the mean of the Beta distribution.
If a site is suitable, these variables should capture factors that would attenuate plant abundance through water availability (i.e., precipitation) or amount plants could grow in winter (proxied by the minimum winter temperature). All covariates were centered and scaled.
To center the prior for the intercept in the degenerate probability, for various intercept values we fit the model to 130  points and computed the average $CRPS_f$ on the 50 held-out points. Additionally, at this stage, we fix $a=1$, thus \textit{a priori} giving equal support above and below 0. We present the models with the lowest $CRPS_f$; centering $\gamma_0$ at different values does not impact significance nor sign of the remaining coefficients.

Table \ref{tab:post_summary_dat} displays posterior means and 95\% credible intervals for both plant families. For Crassulaceae, the large negative $\gamma_0$ intercept (-1.345) suggests that fewer zeros arose from unsuitability; most sites were climatically suitable. The negative coeﬃcient for mean annual potential evaporation indicates that for fixed rainfall concentration, higher evaporation increases the suitability of a location, reasonable given that Crassulaceae are succulent plants adapted for drier conditions. The effect of rainfall concentration on unsuitability was insignificant. Higher annual mean precipitation and colder winter temperatures both tended to favor higher Crassulaceae percent cover, holding the other covariate constant. The Restionaceae typically had results opposite to those of Crassulaceae, and there was evidence that 
more concentrated rainfall throughout the year at a site led to higher suitability compared to sites with similar evaporation. 

We hypothesize that opposing results between the two families are driven by both fire dynamics and subregion-specific climate influences. While members of the Restionaceae can inhabit a wide range of moisture availability levels, restio-dominated fynbos primarily occur on warmer north-facing slopes on drought-prone soils \citep{rebelo_fynbos_2006}. The Restionaceae are also a fire-adapted family, e.g., members germinate in response to smoke \citep{brown_stimulation_1994} and have reseeding or resprouting life cycles \citep{wuest_resprouter_2016}, that often “carry” fires in fynbos \citep{cowling_flora_1992}. Our results agreed with these generalizations in that higher Restionaceae abundances were associated with lower rainfall and higher winter temperatures. Within the Baviaanskloof subregion, an additional dynamic of vegetation turnover may play a role in the Restionaceae suitability patterns we observed. Grassy fynbos, i.e., fynbos dominated by C4 grasses, begins to outcompete restio-dominated fynbos in the eastern cape, particularly on drier north facing slopes. In this context, fynbos tends to occur in wetter areas \citep{rebelo_fynbos_2006}.  This may explain why we observed lower suitability for areas with higher potential evaporation, despite the favor of drier sites for abundance. On the other hand, Crassulaceae are likely “fire avoiders” and may have higher abundances in rockier areas that act as outcropping island microsites \citep{cousins_beating_2016}. Higher Crassulaceae abundance associated with wetter, cooler sites may also reflect a broader signal of lower fire frequency in these areas.

\begin{table}[ht]
\caption{Posterior means and 95\% credible intervals for coefficients from fitting various zero-inflated Beta models to CFR data. For all the following, $c$ was fixed at $0.5$. The $\gamma$ coefficients are for probability of unsuitability, the $\delta$ and $\psi$ coefficients are for the Beta regression. Dashes correspond to intervals that included 0.}
\centering
\begin{subtable}[h]{0.9\textwidth}
\centering
    \caption{Posterior summaries from non-spatial model $\mathcal{M}_1$ fit independently for each species across both Baviaanskloof and Cape Point. Priors for $\gamma_0$ for both species are centered at $-0.75$.}
    \begin{tabular}{|l|ccc|ccc|}
      \hline
      & \multicolumn{3}{c | }{Crassulaceae} &   \multicolumn{3}{c | }{Restionaceae} \\ \hline
     & mean & 2.5\% & 97.5\% & mean & 2.5\% & 97.5\% \\ 
      \hline
    $\gamma_0$: Intercept & -1.345 & -1.765 & -0.876 & -1.0058 & -1.2453 & -0.7412 \\ 
      $\gamma_{1}$: mean\_ann\_pan & -1.269 & -1.960 & -0.377 & 0.884 & 0.294 & 1.630 \\
     $\gamma_{2}$: rainfall\_conc & - & - & - & -0.676 & -1.210 & -0.015 \\ \hline
     $\delta_{0}$: Intercept & 0.238 & 0.041 & 0.473 & 1.685 & 1.420 & 2.115 \\ 
      $\delta_{1}$: mean\_ann\_precip & 0.800 & 0.625 & 0.998 & -0.896 & -1.233 & -0.505 \\ 
       $\delta_{2}$: min\_temp\_july & -0.396 & -0.694 & -0.061 & 0.419 & 0.269 & 0.583 \\  
      $\psi$ & 1.023 & 0.790 & 1.306 & 1.360 & 1.036 & 1.776 \\
       \hline
    \end{tabular}
    \label{tab:post_summary_dat}
    \end{subtable} 
\begin{subtable}[h]{0.9\textwidth}
\centering
    \vspace{0.5cm}
    \caption{Posterior summaries from the spatial zero-inflated Beta models $\mathcal{M}_2$ and $\mathcal{M}_3$ fit to the Crassulaceae percent cover data in Baviaanskloof. $\mathcal{M}_2$ includes the spatial effect in $\mu_i$, with the prior for $\delta_0$ centered at 1. $\mathcal{M}_3$ includes the effect in $\pi_i$, with the prior for $\gamma_{0}$  centered at -0.75. For both models, $\phi$ fixed at one-third the maximum observed distance (14.8 km).}
\begin{tabular}{|l|ccc|ccc|}
  \hline
&  \multicolumn{3}{c|}{$\mathcal{M}_2$} &  \multicolumn{3}{c|}{$\mathcal{M}_3$} \\ \hline
 & mean & 2.5\% & 97.5\% & mean & 2.5\% & 97.5\% \\ 
  \hline
$\gamma_{0}$: Intercept & -0.951  & -1.326 & -0.572  & -1.080 & -1.642 & -0.377 \\ 
  $\gamma_{1}$: mean\_ann\_pan & -0.787 & -1.395 & -0.147 & - & - & - \\ \hline
  $\delta_{0}$: Intercept & 0.956 & 0.658 & 1.271 & 0.574 & 0.312 & 0.855 \\ 
  $\delta_{1}$: mean\_ann\_precip & 1.222 & 0.903 & 1.677 & 1.140 & 0.819 & 1.504 \\ 
  $\delta_{2}$: min\_temp\_july & - & - & - & -0.405 & -0.683 & -0.056 \\
$\psi$ & 1.628 & 1.230 & 2.041  & 0.920 & 0.647 & 1.292 \\ \hline
  $\sigma^{2}$ & 1.510 & 1.102 & 2.047  & 0.920 & 0.647 & 1.292 \\ 
   \hline
\end{tabular}
\label{tab:post_sum_spatial}
    \end{subtable}
    \label{tab:post_sum_all}
\end{table}

\subsection{$\mathcal{M}_{2}$ and $\mathcal{M}_{3}$: Results for a subset of the CFR data}
\label{sec:spatial_cfr_analysis}

Baviaanskloof and Cape Point are disjoint and too far apart geographically to envision a common spatial process for the random effects.  So we perform spatial analysis only on Baviaanskloof as two-thirds ($n = $ 119) of the total observations are located in this subregion. 
The analyses presented here are for Crassulaceae; similar analyses for Restionaceae are provided in the Supplement (Tables S\ref{tab:restio_spatial_postsum}, S\ref{tab:rest_mods} and Fig.  S\ref{fig:restio_res}).

\begin{figure}[ht]
    \caption{Posterior means of spatial random effects under the two spatial models $\mathcal{M}_2$ (left) and $\mathcal{M}_3$ (right) fit on the {Crassulaceae} data in Baviaanskloof. Please see Table S\ref{tab:crass_mods}}
    \centering
    \includegraphics[scale = 0.4]{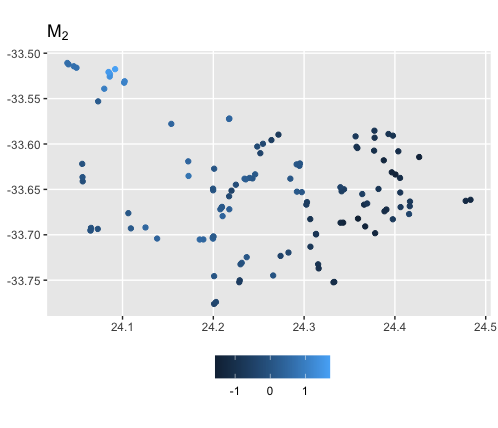} \quad
    \includegraphics[scale = 0.4]{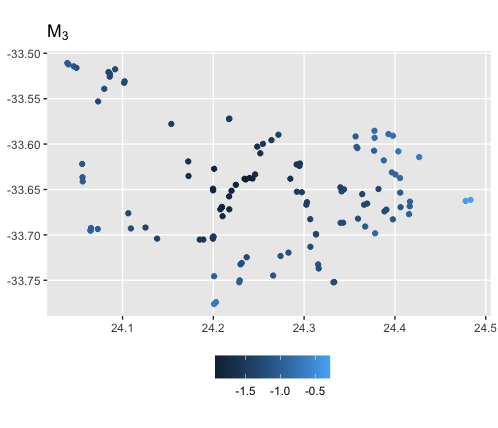}
    \label{fig:crass_res}
\end{figure}

In fitting the two spatial models to the Crassulaceae data in Baviaanskloof, we fixed $a=1$ and experimented with different informative priors for the respective intercepts. Similar to Section \ref{sec:nonspatial_analysis}, we ultimately select a prior which leads to the best $CRPS_{f}$. We began with the same covariates used to fit the non-spatial $\mathcal{M}_1$ in Section \ref{sec:nonspatial_analysis}. When fitting $\mathcal{M}_2$, minimum July temperature was no longer significant for $\mu(\bs_i)$. For $\mathcal{M}_{3}$, average annual evaporation was no longer significant for $\pi(\bs_i)$. 
 Otherwise, the ecological results and interpretations found in Section 5 still hold (Table \ref{tab:post_sum_spatial}). 
The estimated spatial variance under $\mathcal{M}_{2}$ is larger than that of $\mathcal{M}_{3}$ (see Fig. \ref{fig:crass_res} for estimates of the random effects).

\subsection{Model comparison for the CFR data}
\label{sec:spatial_comp}
Lastly, to compare the spatial and non-spatial models, we randomly split the data into a training set of 99 locations and a test set of 20 locations, and fit five models--BEZI, $\mathcal{M}_0$, $\mathcal{M}_1$, $\mathcal{M}_2$, and $\mathcal{M}_3$--using the covariates that were found significant for each respective model. Additionally, we fix $c$ at three different values: $\{0.1, 0.3, 0.5\}$ for all of the $\mathcal{M}$ models in order to assess whether the data are sensitive to the choice of $c$. We compare the models using the test data by evaluating Tjur's $R^2$ and AUC for all five models, and the two $CRPS$ metrics for $\mathcal{M}_1$-$\mathcal{M}_3$. We repeat this process thirty times and average the results. Table S\ref{tab:crass_mods} shows that the model with the spatial effect in the mean, $\mathcal{M}_2$, with $a=1$ outperforms the other four models in all the comparison metrics. This could be related to the estimated larger spatial variance under $\mathcal{M}_2$. Changing $c$ leads to minor differences in predictive performance. 
The BEZI model exhibits the lowest performance in classifying between zero and positive percent covers for the data (lowest $R^2$ and AUC, Table S\ref{tab:crass_mods}). This suggests that modeling two sources of zeros is more appropriate for the data, with the inclusion of spatial dependence for $\mu_{i}$ yielding the best fit. Comparisons for Restionaceae are presented in Table S\ref{tab:rest_mods}.

\section{Conclusions}

Modelling zero-inflated percent cover data has been a statistical challenge for ecologists.  While many approaches have been proposed, we have argued that these do not satisfactorily address the zero-inflation issue. We have developed a left-censored approach to obtain zeros for data on the unit interval, and offered a zero-inflated model for percent or proportion data which parallels such modeling for count data. The specifications enable better understanding of the incidence of absence as well as the explanatory contribution of environmental unsuitability versus absence by chance. Given that ecological data are often spatial, we have supplied spatial models to enhance explanation and enable prediction to unobserved sites with known environmental attributes. Better ecological insights emerge, particularly illustrated by our example with plant percent cover data.

Future work, in progress, considers joint zero-inflation. For example, we can examine percent cover for a pair of species at a site to allow deeper focus on biotic factors (i.e., interspecific competitive interactions) that influence site suitability and relative species abundances. 
Given the sum constraint at a site to at most 1, this framework anticipates negative association between pairs of species.

\section*{Appendix}

Here, we offer analytical insight into the effect of the choice of support for the latent variable, $W$ (suppressing the site subscript). 
By letting $W_{a} = -a + (1+a)V$ where $V \sim Beta(\mu, \nu)$, $W$ has support $[-a,1)$. Again, $a=1$ provides matching support above and below $0$.  Then, $P(W_{a} \leq 0) = P(V \leq \frac{a}{a+1} \equiv c(a))$.  
Though $W_a$ is a linear transformation of $W_1$, there is no linear transformation of the Beta distribution that moves $c$ to $.5$, corresponding to $a=1$.  We cannot ``relocate'' $\mu$ for general $a$ to $\mu$ for $a=1$.  The shape of the Beta changes with $\mu$.

It is more convenient to work with $c$ since we seek the effect of changing $\mu$ for a fixed $\nu$ and $c$ is on the support of $V$.  We seek to understand the behavior of $P(V \leq c|\mu, \nu)$.
First, for a fixed value $p$ of this probability and also a fixed $\nu$, qualitatively, $c \uparrow \mu$.  Pursuing this analytically, consider the implicit function, $F(c_{p,\nu}(\mu); \mu, \nu) = p$ where $F$ denotes a Beta cdf.
Plots of the function, $c_{p,\nu}(\mu)$ vs. $\mu$, from $(0,1)$ to $(0,1)$ appear sigmoidal, i.e., with asymptotes at $0$ and at $1$ (see Figure S\ref{fig:plots_mu_c}).

$S$-shaped curves are often characterized by an ordinary differential equation.  For example, the generalized logistic functions (Richards' curves) are a rich class including the Gompertz and Weibull with such an attractive characterization as well as an explicit solution (a convenient current cite is $https://en.wikipedia.org/wiki/Generalisedlogisticfunction$).  Here, we produce the differential equation  which $c_{p,\nu}(\mu)$ satisfies. It is not an ordinary differential equation and, in general, it has no closed form solution since $F(\cdot; \mu, \nu)$ has no closed form. We have $p= F(c_{p,\nu}(\mu); \mu, \nu) = \int_{0}^{c_{p,\nu}(\mu)} f(x; \mu, \nu)dx$ where $f$ is the Beta$(\mu, \nu) $ density.  Using Leibniz's rule, we take the derivative of both sides yielding
$$ f(c_{p,\nu}(\mu); \mu, \nu) dc_{p, \nu}(\mu))/d\mu + \int_{0}^{c_{p, \nu}(\mu)} \frac{\partial f(x; \mu, \nu)}{\partial x} dx =0.$$
Solving, we have the differential equation

\begin{equation}
dc_{p, \nu}(\mu))/d\mu = \frac{ -\int_{0}^{c_{p, \nu}(\mu)} \frac{\partial f(x; \mu, \nu)}{\partial x} dx}{f(c_{p,\nu}(\mu); \mu, \nu)}.
\label{eq:DE}
\end{equation}

Restoring the $(\alpha,\beta)$ notation, the integral in the numerator exists if $\alpha >1$ and $\beta >1$ (implying $\nu >2$); otherwise $c_{p,\nu}(\mu)$ is not differentiable.  If the integral exists, we have
\begin{equation}
 -\int_{0}^{c_{p, \nu}(\mu)} \frac{\partial f(x; \mu, \nu)}{\partial x} dx = (\alpha + \beta +1)[F(c_{p,\nu}(\mu); \alpha, \beta-1) - F(c_{p,\nu}(\mu); \alpha - 1, \beta)].
 \label{eq:DEmore}
 \end{equation}
Clearly this expression is $>0$ and since the denominator is positive, the monotone increasing behavior of $c_{p, \nu}(\mu)$ is demonstrated.  The support for the existence over $(\nu, \mu) \in (0, \infty) \times (0,1)$ satisfies $\mu >  1/\nu$ and $\mu < 1- 1/\nu$.

For $c_{p, \nu}(\mu)$ to be $S$-shaped, we need to show that it has exactly one point of inflection, i.e., that $dc_{p, \nu}(\mu))/d\mu $ increases up to a point and then decreases.  Again using Leibniz's rule, we can take the derivative of $[F(c_{p,\nu}(\mu); \alpha, \beta-1) - F(c_{p,\nu}(\mu); \alpha - 1, \beta)]$.  This derivative exists if $\alpha >2$ and $\beta >2$.  So, now $\nu > 4$ and the support for the existence of the second derivative is $\mu >  2/\nu$ and $\mu < 1- 2/\nu$.  Omitting details, in special cases we can demonstrate that this derivative starts positive and then becomes negative, implying a unique point of inflection.

Finally, to return to sensitivity to choice of $c$, we fit $S-$shaped curves to $c_{p, \nu}(\mu)$ for various $p$ and $\nu >4$.  In the literature, the support of the $S$-shaped curves is always on $R^1$ while we need support  $(0,1)$.  We propose to use the logit function $g_{1}(x) = \text{ln}\frac{x}{1-x}$ from $(0,1)$ to $R^1$.  Then, we apply the generalized logistic function, $g_{2}(t) = (1 + \beta_{0}e^{-\beta_{1}t})^{-1/\gamma}$.  
Plugging in, we obtain the class $g_{2}(g_{1}(x)) = \left(1+ \beta_{0}\left(\frac{1-x}{x}\right)^{\beta_{1}}\right)^{-1/\gamma}$.
This three parameter function is straightforward to fit to $c_{p,\nu}(\mu)$. Implementing the curve fitting over a range of $p$ and $\nu$ (Figure S\ref{fig:curve_fit}), we see that the curves are essentially indistinguishable implying that there is little sensitivity to choice of $a$.

\clearpage
\bibliography{references}

\clearpage
\title{Supplementary Information}
\maketitle
\setcounter{figure}{0}  
\setcounter{table}{0}  
\setcounter{section}{0}
\setcounter{equation}{0}  

\section{Monte Carlo integration for CRPS}
\label{sec:mc_integration}

For observation $y_i$ and given $N$ uniform samples $x^{(j)}$, \begin{equation}
    \int_{0}^{\infty} (F(x) - \bone( y  < x ))^2 dx \approx \frac{1}{N} \sum_{j=1}^{N} \left({F}(x^{(j)}; \alpha_{i}, \beta_{i})  - \bone( y_i  < x^{(j)} )\right)^2
\end{equation}

If the posterior predictive CDF $F$ does not have a closed form, then given a sequence of $\{\theta_{j}\}_{j=1}^{m}$ of parameter values from the posterior distribution, $F$ can be approximated via $\hat{F}(x) \approx \frac{1}{m} \sum_{j=1}^{m} F_c(x | \theta_j)$, where $F_c(x|\theta)$ is the conditional predictive CDF.

\clearpage
\section{Simulation details for comparing BEZI, LCEB, and zero-inflated Beta models}
\label{sec:simulation_details}
In Section 4.2 of the main text, we present results of simulations to examine the prediction performances of the BEZI model of \cite{ospina2010inflated}, our censor-only model LCEB, and our zero-inflated Beta model $\mathcal{M}_{1}$. In all cases, we generate $n_{test}= 100$ and $n_{train} = 200$ observations from one model, fit all three models to the train data, and obtain predictions for the test data. We compare the predictions to the true data by calculating $R^2$ and AUC for the probabilities of classifying on observation as 0 or positive, and average $CRPS_{h}$ for all positive test observations. For all simulations, we set the true $\nu = 4.5$, i.e. just an intercept for the regression for $\nu$ (see the appendix for this choice of $\nu$). Chains were run for 5000 iterations after 2500 burnin, and thinned to retain every fifth sample. For models LCEB and $\mathcal{M}_{1}$, the data are generated and models fit with $a= 1$. For all the parameters in all three models, we use diffuse $N(0, 10)$ priors. The only exception is that we employ an informative prior for $\gamma_{0}$ when fitting $\mathcal{M}_{1}$, as described below.  

For data generated from the zero-inflated Beta model $\mathcal{M}_{1}$, the regressions for $\pi_{i}$ and $\mu_{i}$ each contain an intercept and one independent $N(0,1)$ covariate, i.e. $Z_{i} = (1 \ Z_{i1})'$ and $X_{i} = (1 \ X_{i1})'$, where $Z_{i1} \sim N(0,1)$ independent of $X_{i1} \sim N(0,1)$.  The proportions of 0s and their sources are controlled by varying the intercepts $\gamma_{0}$ and $\delta_{0}$. The BEZI model is fit with the same $\bZ$ and $\bX$ for the regressions for $\pi$ and $\mu$, respectively. LCEB is fit by concatenating or joining the two covariate sets together, i.e. $X_{i,LCEB} = (1 \ X_{i1} \ Z_{i1})'$. When fitting $\mathcal{M}_{1}$, we use a $N(\gamma_{0, true}, \sigma^2 = 0.25)$ prior for $\gamma_{0}$.

For data generated from the BEZI model, the regressions for $\pi_{i}$ and $\mu_{i}$ once again each contain one independent $N(0,1)$ covariate. The proportion of 0s is controlled by varying the intercept $\gamma_{0}$, and the magnitude of the positive percent covers is controlled by $\delta_{0}$. $\mathcal{M}_{1}$ is fit with the same $\bZ$ and $\bX$ for the regressions for $\pi$ and $\mu$, respectively. LCEB is fit by concatenating or joining the two covariate sets together, i.e. $X_{i,LCEB} = (1 \ X_{i1} \ Z_{i1})'$. When fitting $\mathcal{M}_{1}$, we use a $N(\gamma_{0, true}, \sigma^2 = 0.25)$ prior for $\gamma_{0}$.

For data generated from the censor-only model LCEB, we only have a single covariate set $\bX$, where $X_{i} = (1 \ X_{i1})'$ and $X_{i1} \sim N(0,1)$. The BEZI and $\mathcal{M}_{1}$ models are fit with $\bZ = \bX$. When fitting $\mathcal{M}_{1}$, we use a $N(0, \sigma^2 = 0.25)$ prior for $\gamma_{0}$.

\clearpage 
\section{Posterior distributions for spatial zero-inflated Beta}
\label{sec:spatial_posts}
If we place the spatial random effects $\bet(\bs)$ into the mean $\mu(\bs_i)$ and hold the spatial decay parameter $\phi_{\eta}$ fixed, then the posterior distribution is proportional to:
\begin{equation}\label{eq:posterior_re1}
\prod_{i=1}^{n} f_{ {Y_i} | Z_i, \bdelta, \bpsi} (Y_i | Z_i, \bdelta,\bpsi, \eta(\bs_i) )  \cdot \prod_{i=1}^n f_{Z_i | \bgamma}(Z_i | \bgamma) \cdot 
 f_{\bet | \sigma^{2}}(\bet(\bs) | \sigma^{2}_{\eta} ) \cdot f_{\bgamma}(\bgamma) \cdot f_{\bdelta}(\bdelta)  \cdot f_{\bpsi}(\bpsi)\cdot f_{\sigma^{2}}(\sigma^{2}_{\eta}).
\end{equation}

We could also use a Gaussian process for the spatial effects $\btau(\bs)$ in $\pi(\bs_i)$. Once again holding the spatial decay parameter $\phi_{\tau}$ fixed, the posterior is proportional to:
\begin{equation}\label{eq:posterior_re2}
\prod_{i =1}^{n} f_{Y_i | Z_i, \bdelta, \bpsi} (Y_i | Z_i, \bdelta,\bpsi) \cdot \prod_{i=1}^n f_{Z_i | \bgamma, \tau(\bs_i)}(Z_i | \bgamma, \tau(\bs_i)) \cdot  f_{\btau | \sigma^{2}}(\btau(\bs) | \sigma^{2}_{\tau} ) \cdot f_{\bgamma}(\bgamma) \cdot f_{\bdelta}(\bdelta)  \cdot f_{\bpsi}(\bpsi)\cdot f_{\sigma^{2}}(\sigma^{2}_{\tau}).
\end{equation}

\clearpage
\section{Supplementary figures}
\begin{figure}[h]
    \caption{Histograms of environmental covariates within Baviaanskloof (grey) and Cape Point (green).}
    \centering
    \includegraphics[scale = 0.35]{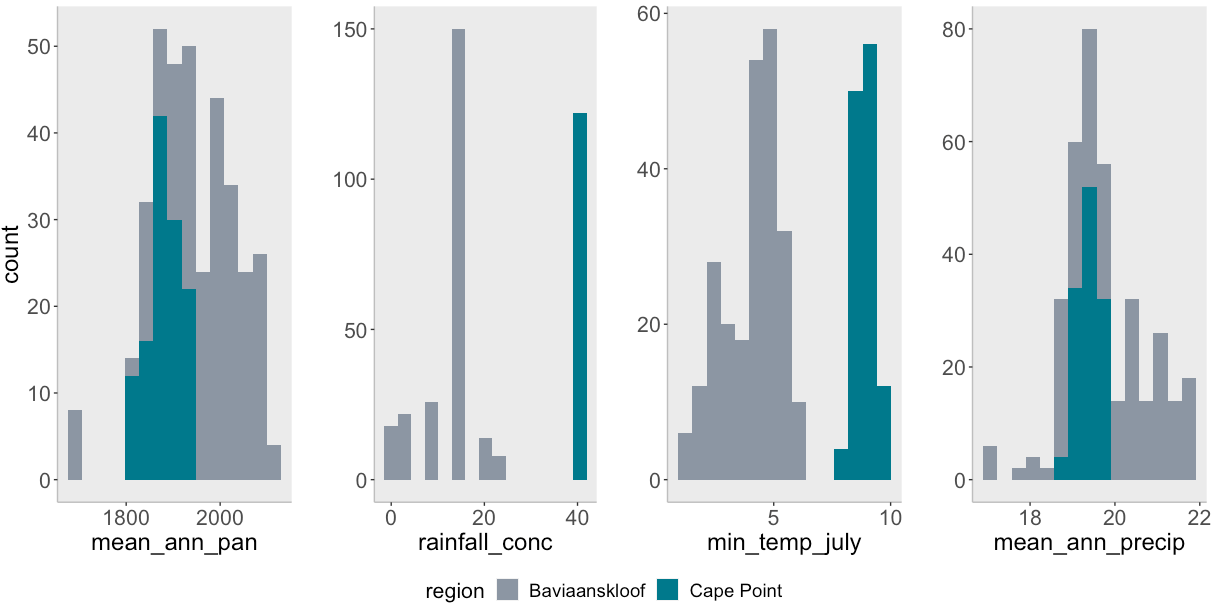}
    \label{fig:eda_covars}
\end{figure}

\begin{figure}[h]
    \caption{Spatial plots of percent cover by species in Cape Point. Orange points denote observed 0, and blue points are positive covers.}
    \centering
    \includegraphics[scale = 0.5]{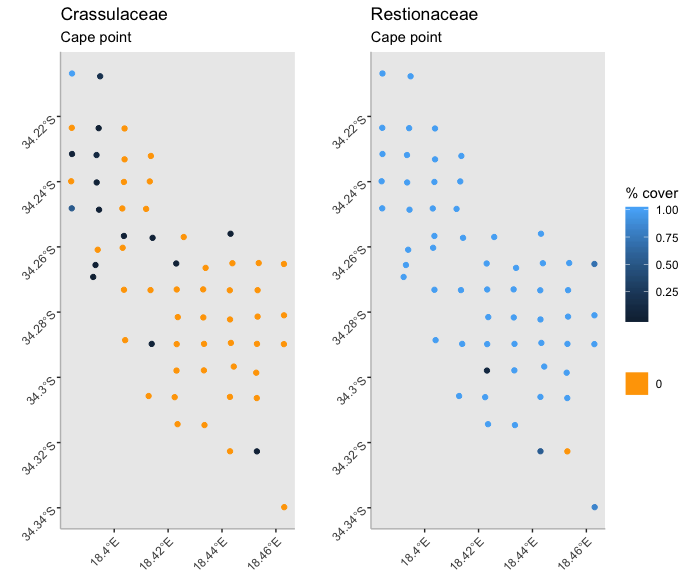}
    \label{fig:spatial_cp_pcts}
\end{figure}

\begin{figure}[h]
    \caption{Results of simulation study for censored model. Data generated under various values of $c$ (panel titles), then fit using models with $c$ held fixed (x-axis). Regression for $\mu$ has a single covariate, i.e. $logit(\mu) = \delta_{0} + \delta_{1} x_{1}$, with varying $\delta_{0}$ and $\psi$. the average proportion of observed 0s in the training data. For Simulations 1-6: $\delta_{0} = \{-0.5, 0.5, 1.5, -0.5, 0.5, 1.5\}$ and $\phi = \{3,3,3,15,15,15\}$. Proportions of 0 in data increase with increasing true $c$.}
    \centering
    \begin{subfigure}{0.9\textwidth}
    \caption{Empirical credible interval coverage for $\delta_{1} = -0.5$ with nominal 95\% coverage (dashed line). In all simulations, empirical coverage is closest to nominal when $c_{fixed}$ is set equal to the true $c$, as expected.}
     \includegraphics[scale = 0.35]{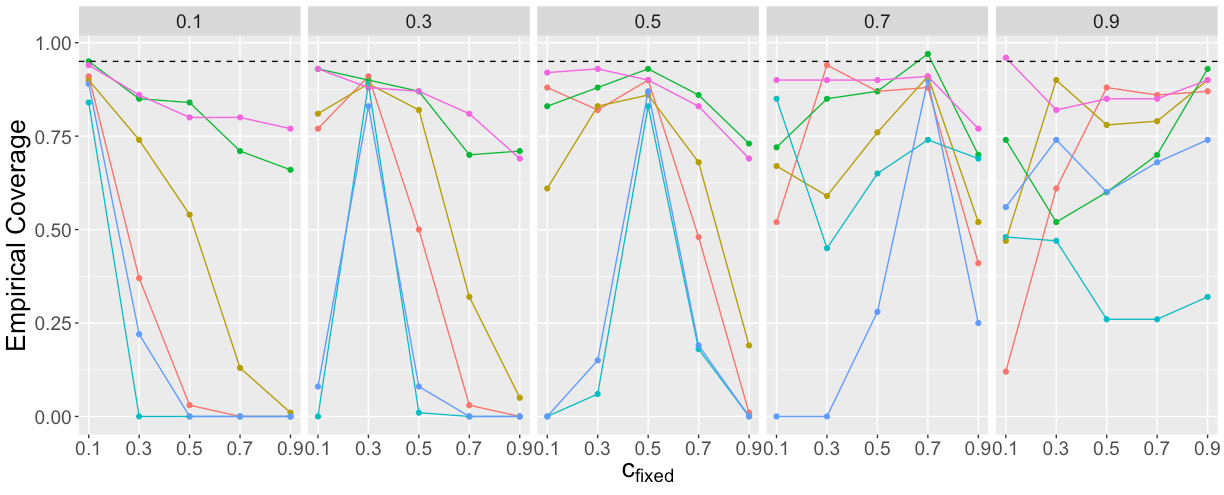}
     \label{fig:emp_cvg_c}
    \end{subfigure}
    \begin{subfigure}{0.9\textwidth}
        \caption{Average RMSPE for the probability of zero for test data. Prediction performance tends to worsen as the magnitude $|c_{fixed} - c_{true}|$ increases (first and last panels).}
     \includegraphics[scale = 0.35]{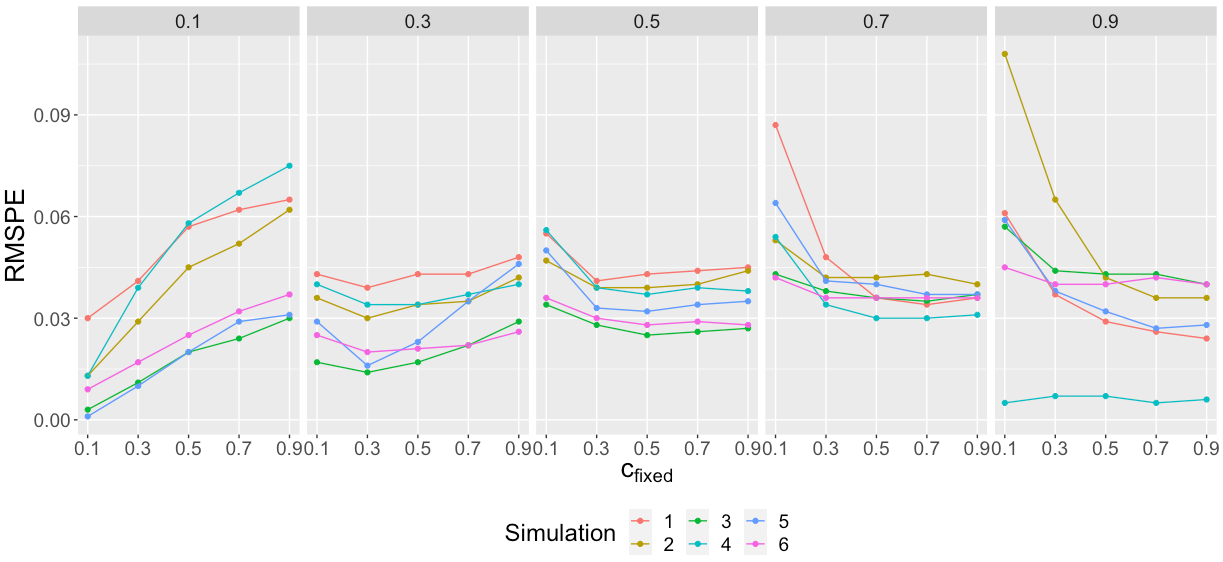}
     \label{fig:rmspe_c}
    \end{subfigure}
    \label{fig:supp_c}
\end{figure}

\begin{figure}
    \centering
    \caption{Plots display the proportion of simulations where sign of $\delta_{1} = -0.5$ was correctly identified according to 95\% credible intervals. Data were generated under various values of $c$ (panel titles), then fit using models with $c$ held fixed (x-axis) with $logit(\mu) = \delta_{0} + \delta_{1} x_{1}$. }
     \includegraphics[scale = 0.35]{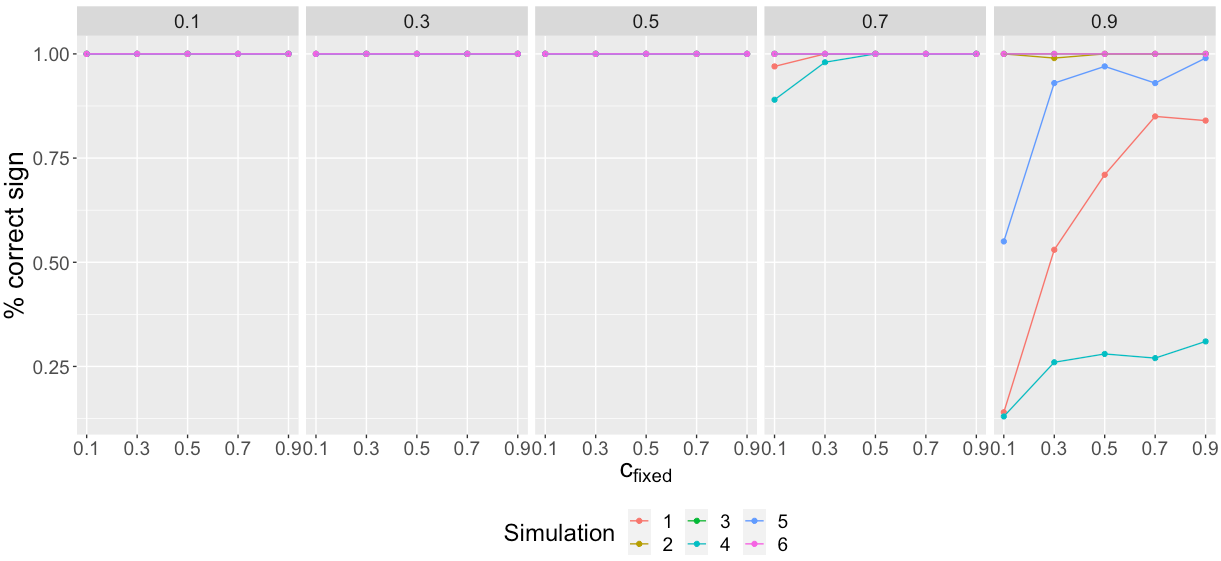}
     \label{fig:prop_sign_c}
\end{figure}

\begin{figure}
    \centering
    \caption{Predicted and true probabilities of 0 for data generated in Simulation 1. Each plot corresponds to data generated with different $c_{\text{true}}$, with colors denoting posterior mean probabilities for models fit with different $c_{\text{fixed}}$. For a given $c_{\text{true}}$, predicted probabilities are generally quite similar across the $c_{\text{fixed}}$. The notable exception is the under-prediction from the model fit with $c_{\text{fixed}} = 0.1$ to data generated with $c_{\text{true}} = 0.9$.}
        \includegraphics[scale = 0.35]{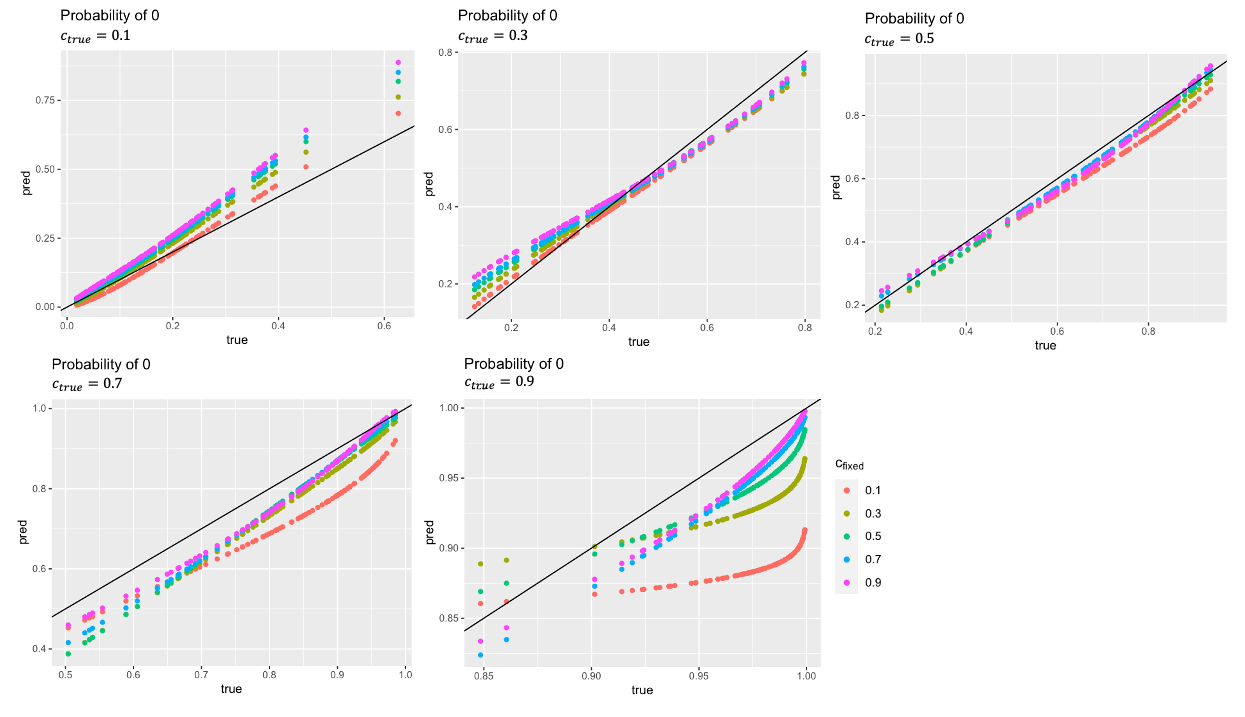}
    \label{fig:censor_sim1_prob0}
\end{figure}

\begin{figure}[h]
    \caption{ROC curves and AUC for Simulations 1.1-1.3 for determining the source of a zero. All curves lie well about the 45-degree line, with AUCs of 0.80 or higher.}
    \centering
    \includegraphics[scale = 0.35]{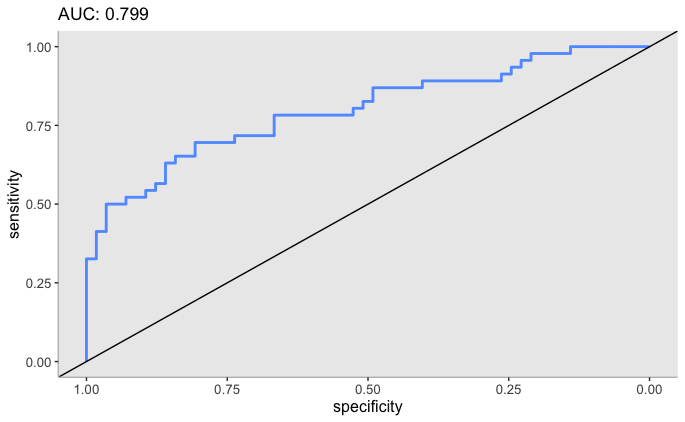}        \includegraphics[scale = 0.35]{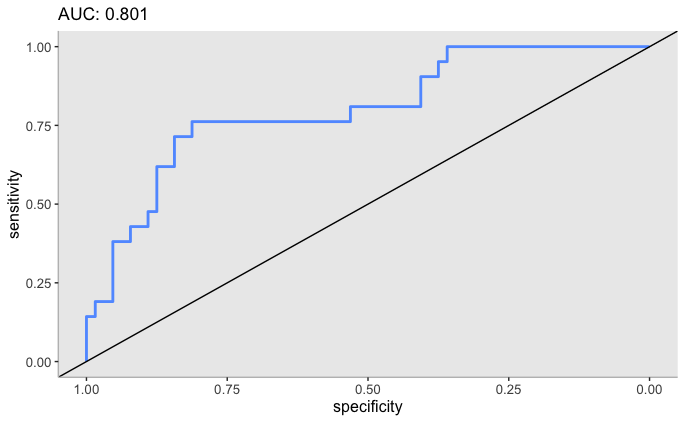}
            \includegraphics[scale = 0.35]{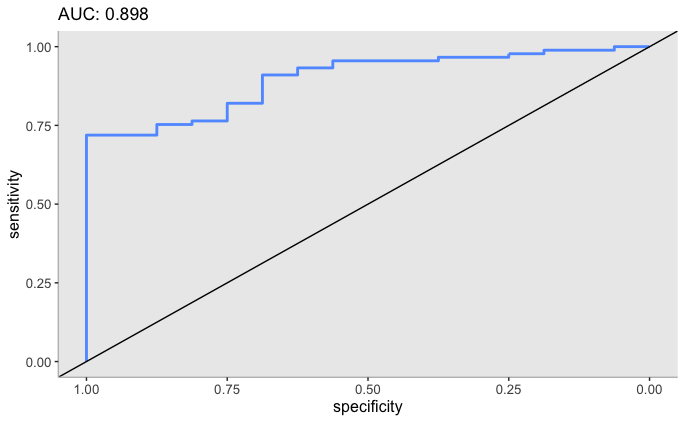}
    \label{fig:m1_roc}
\end{figure}

\begin{figure}
    \caption{Run times (seconds) of zero-inflated Beta models for different sizes $n$ of dataset. $\mathcal{M}_{1}$ refers to the non-spatial zero-inflated Beta model, whereas $\mathcal{M}_{2}$ has space in $\mu(\bs_{i})$ and $\mathcal{M}_{3}$ has space in $\pi(\bs_{i})$. Run times are averaged across data with different proportions and sources of 0s. Simulations are run using a server node with one core Intel(R) Xeon(R) CPU E5-2680 v3 @ 2.50GHz.}
    \centering
    \includegraphics[scale = 0.75]{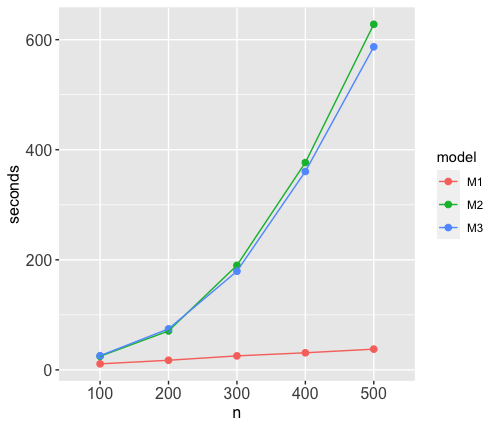}
    \label{fig:run_times}
\end{figure}

\begin{figure}
    \caption{Simulations 2.1-2.3: the first two columns show the true and posterior mean predictions of the spatial random effects, using a common color scale. The third column shows the differences between the true and predicted values. That the majority of the points in the differences plots are white or pale demonstrates that we are able to recover the spatial surface well. See Table S\ref{tab:sims_m2} for corresponding posterior summaries.}
    \centering
    \includegraphics[scale = 0.3]{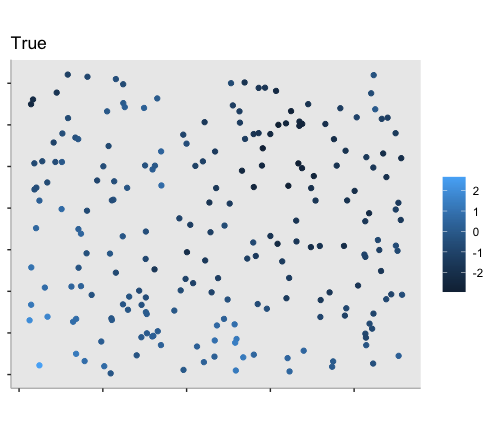}
    \includegraphics[scale = 0.3]{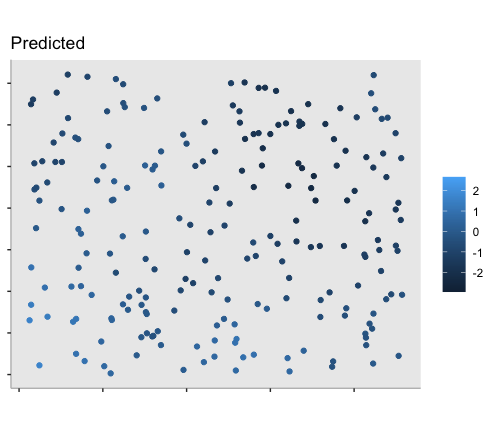}
    \includegraphics[scale = 0.3]{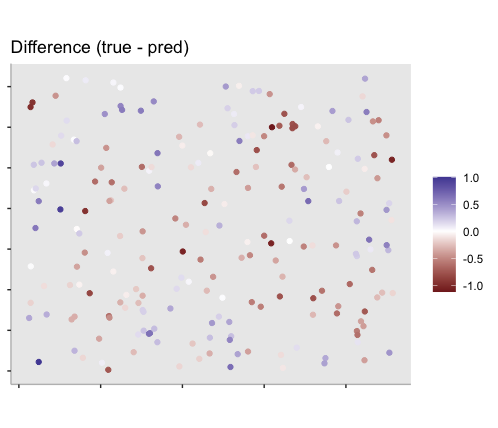}\\
    \includegraphics[scale = 0.3]{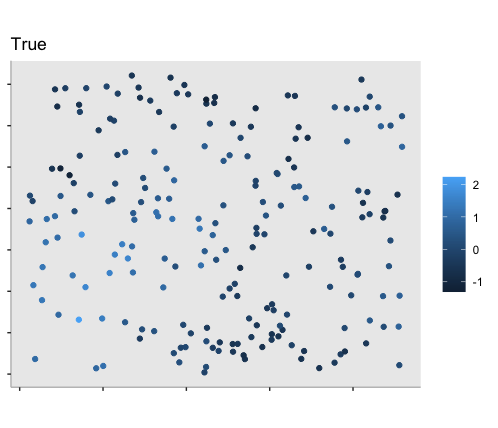}
     \includegraphics[scale = 0.3]{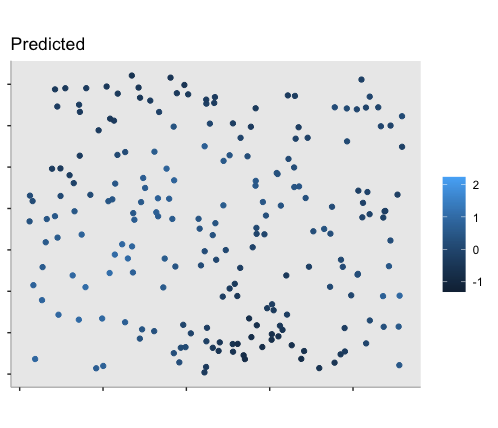}
    \includegraphics[scale = 0.3]{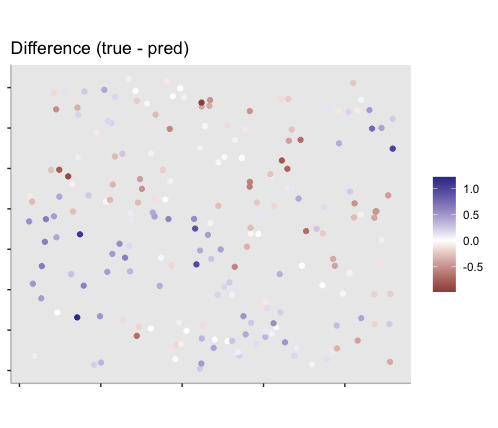}\\
    \includegraphics[scale = 0.3]{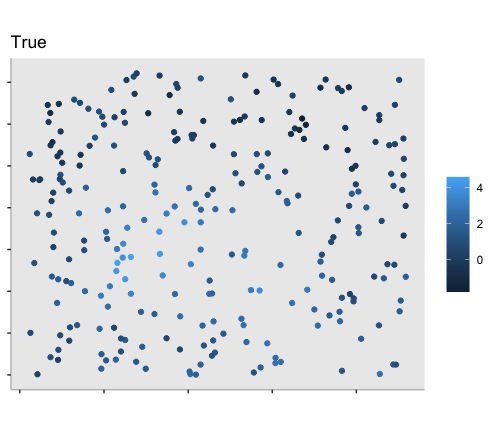}
     \includegraphics[scale = 0.3]{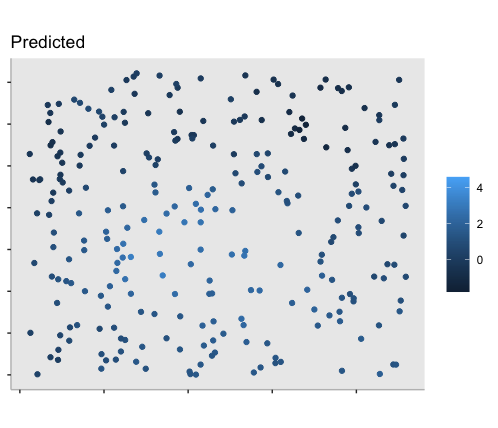}
    \includegraphics[scale = 0.3]{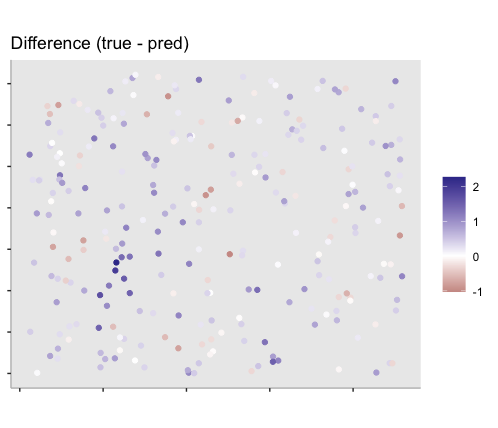}\\
    \label{fig:sim2_re}
\end{figure}

\begin{figure}[h] 
    \caption{Simulations 2.1-2.3: red denotes to the data-generating model $\mathcal{M}_2$, and blue denotes the incorrect model $\mathcal{M}_{1}$. Left: ROC curves for predicting source of zero. Middle: parallel histograms reveal $\mathcal{M}_2$'s superior performance in separating zero and positive observations. Right: Plots of estimates means and probabilities of zero reveal that $\mathcal{M}_{2}$ better approximates the truth when compared to $\mathcal{M}_{1}$.}
    \centering
    \includegraphics[scale = 0.28]{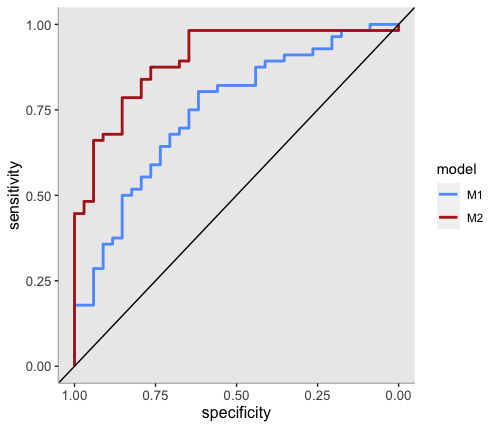}
    \includegraphics[scale = 0.3]{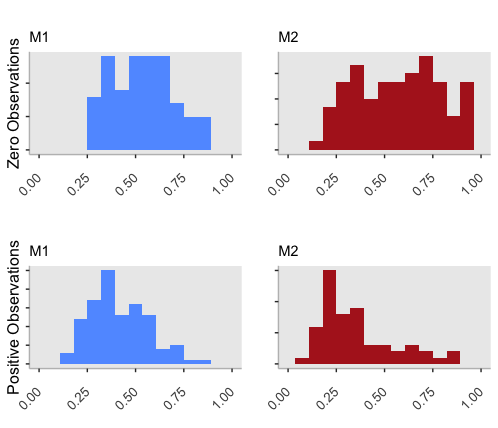}
    \includegraphics[scale = 0.3]{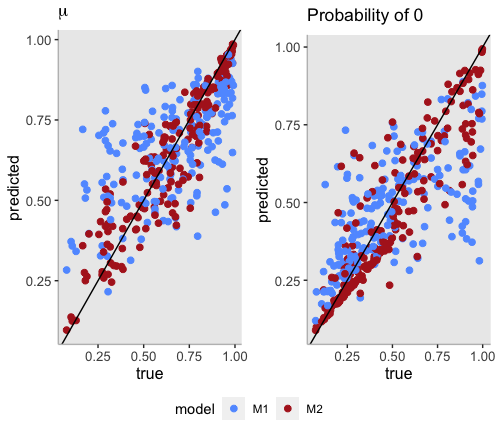}
    \includegraphics[scale = 0.28]{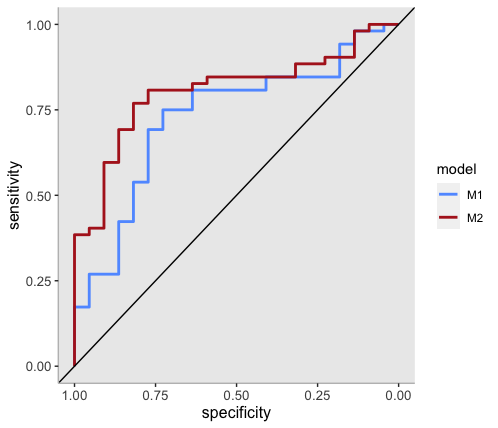}
    \includegraphics[scale = 0.3]{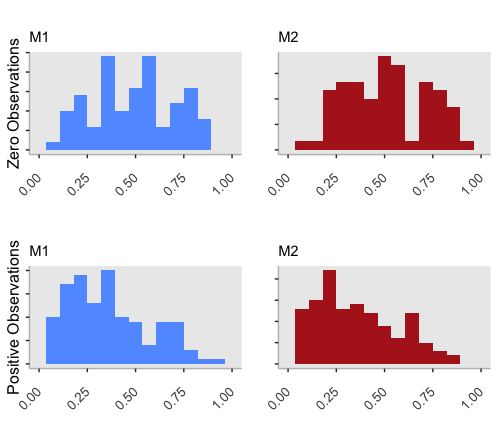}
    \includegraphics[scale = 0.3]{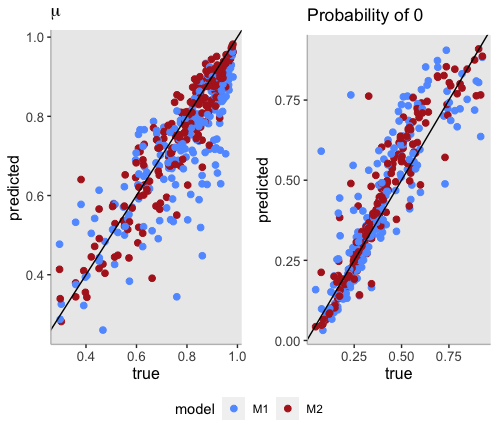}
    \includegraphics[scale = 0.28]{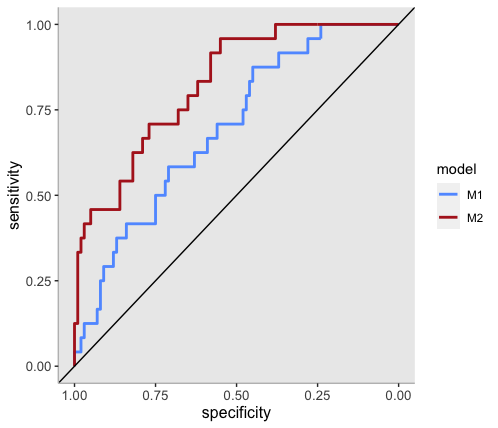}
    \includegraphics[scale = 0.3]{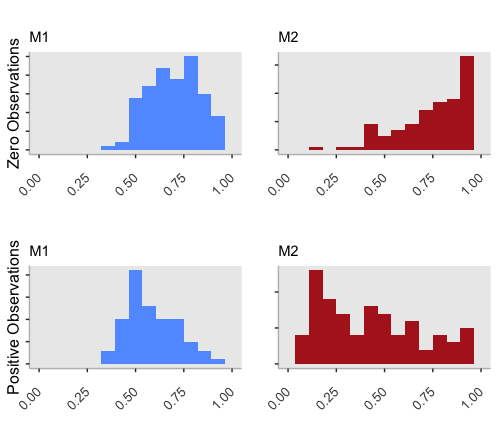}
    \includegraphics[scale = 0.3]{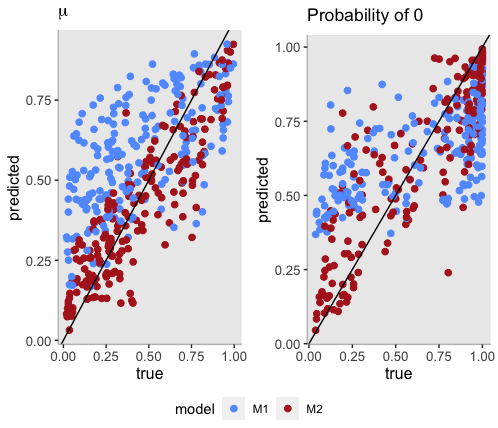}
    \label{fig:re1_comp1} 
\end{figure}

\begin{figure}
    \caption{Simulations 3.1-3.3: the first two columns show the true and posterior mean predictions of the spatial random effects, using a common color scale. The third column shows the differences between the true and predicted values. See Table S\ref{tab:sims_m3} for corresponding posterior summaries. }
    \centering
    \includegraphics[scale = 0.3]{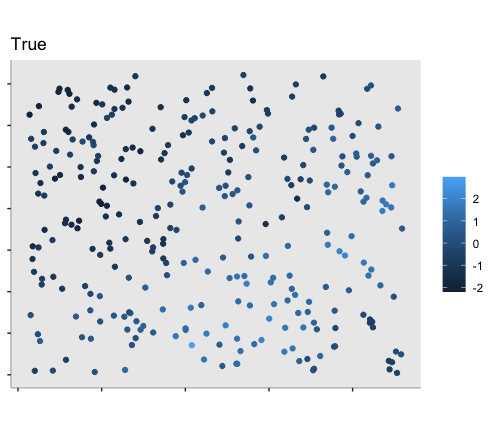}
    \includegraphics[scale = 0.3]{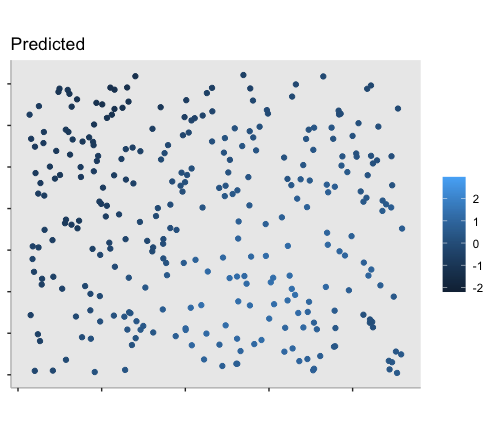}
    \includegraphics[scale = 0.3]{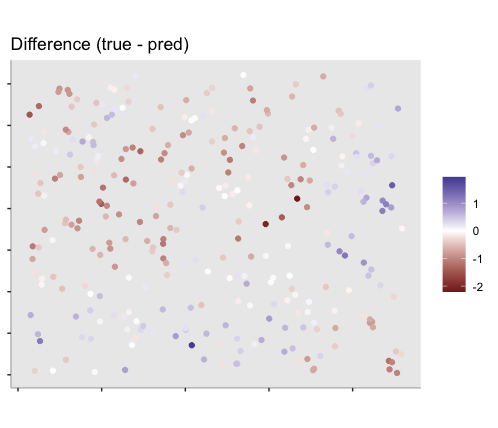}\\
    \includegraphics[scale = 0.3]{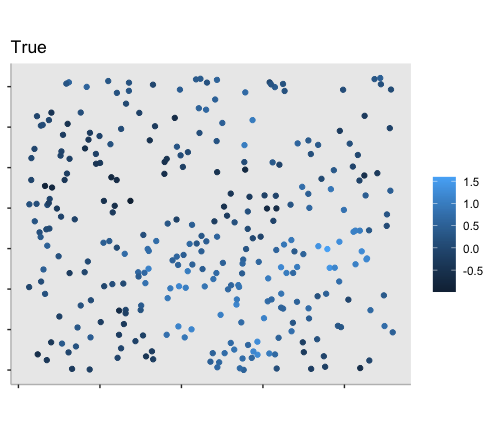}
     \includegraphics[scale = 0.3]{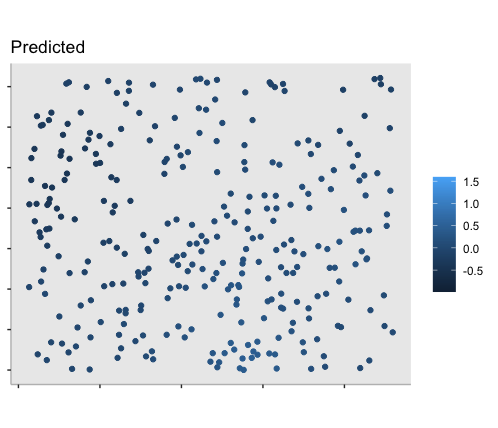}
    \includegraphics[scale = 0.3]{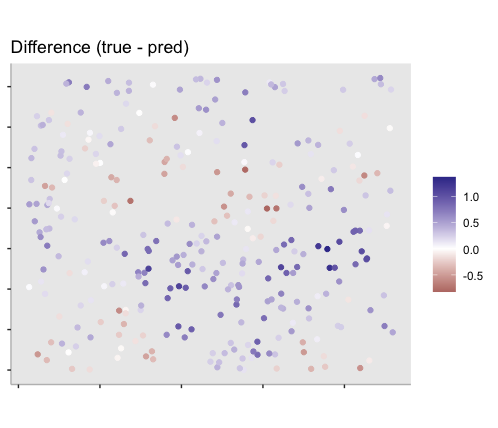}\\
    \includegraphics[scale = 0.3]{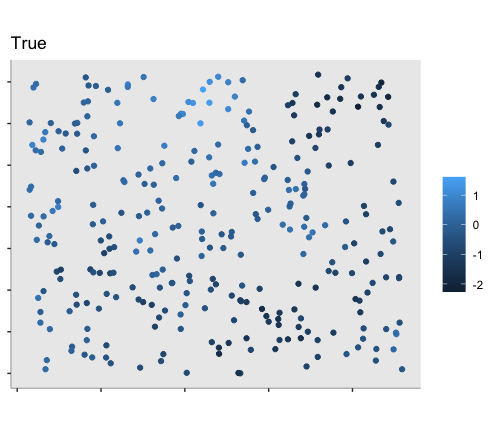}    
    \includegraphics[scale = 0.3]{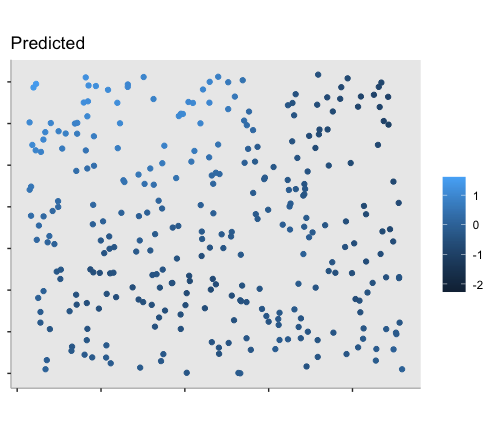}
    \includegraphics[scale = 0.3]{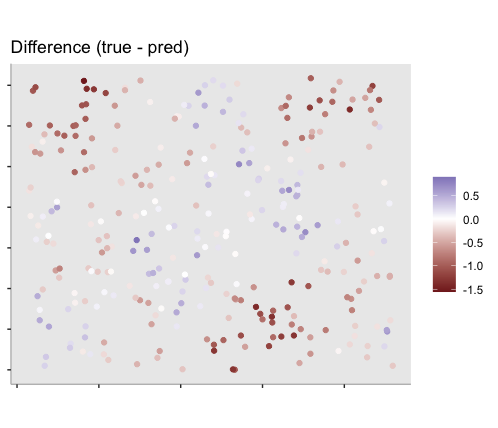}
    \label{fig:sim3_re}
\end{figure}

\begin{figure}[h] %
    \caption{Simulations 3.1-3.3: green denotes to the data-generating model $\mathcal{M}_3$, and blue denotes the incorrect model $\mathcal{M}_{1}$. Left: ROC curves for predicting source of zero. Middle: parallel histograms reveal $\mathcal{M}_3$'s superior performance in separating zero and positive observations. Right: Plots of estimates means and probabilities of zero show that $\mathcal{M}_{3}$ better approximates the true probability of 0 when compared to $\mathcal{M}_{1}$, but both models perform similarly in recovering the true mean.}
    \centering
    \includegraphics[scale = 0.28]{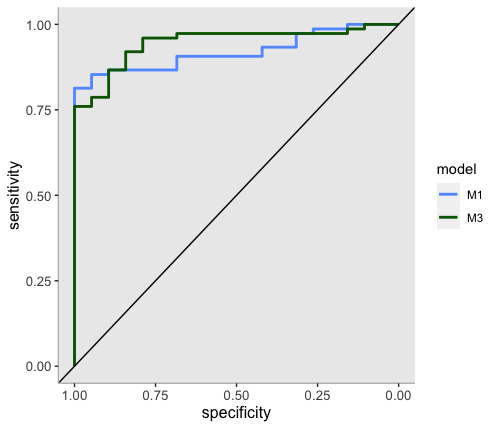}
     \includegraphics[scale = 0.3]{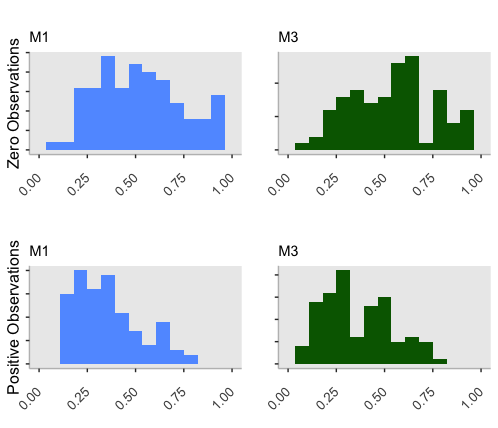}
     \includegraphics[scale = 0.3]{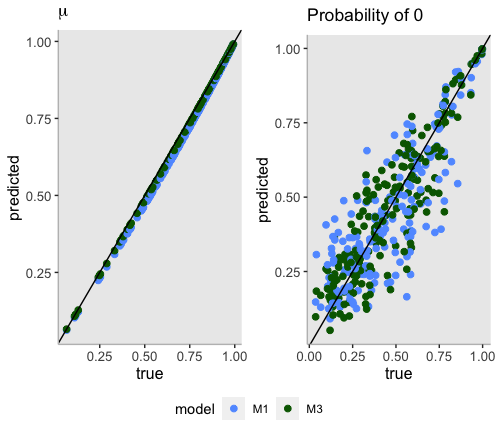}\\
    \includegraphics[scale = 0.28]{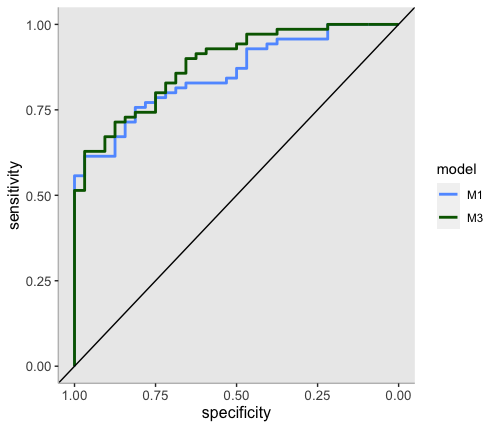}
    \includegraphics[scale = 0.3]{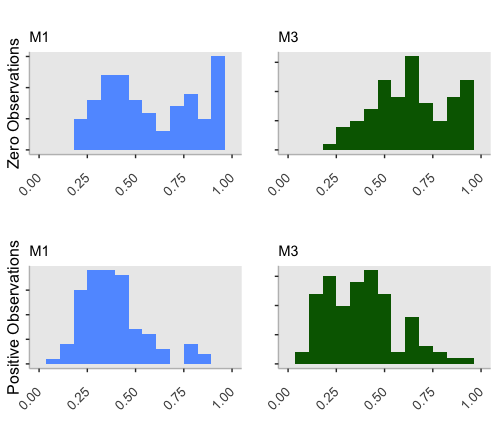}
    \includegraphics[scale = 0.3]{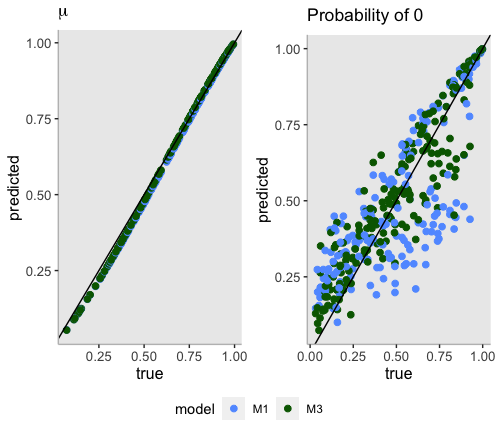}\\
    \includegraphics[scale = 0.28]{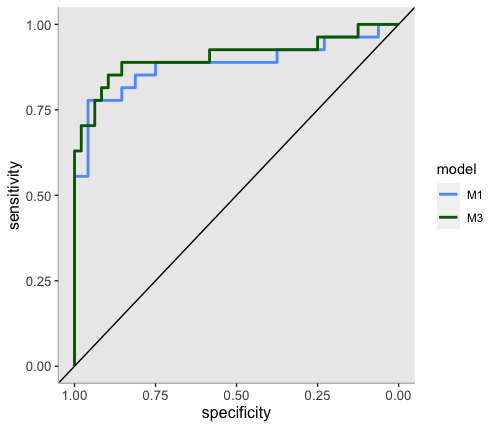}
    \includegraphics[scale = 0.3]{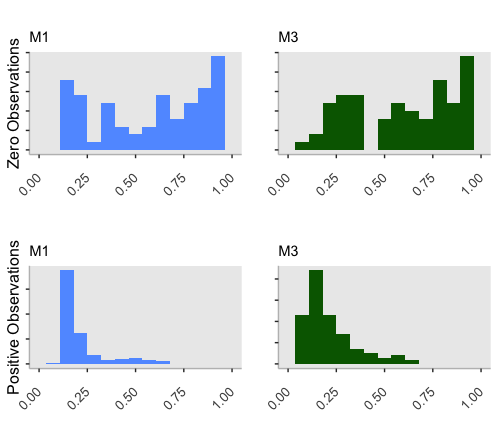}
    \includegraphics[scale = 0.3]{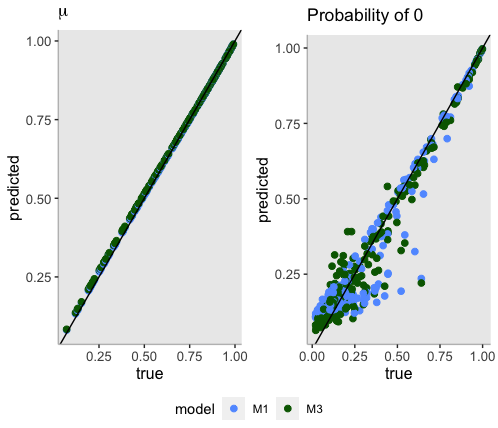}
    \label{fig:re2_comp_plots} 
\end{figure}

\begin{figure}[h]
    \caption{Posterior means of spatial random effects under the two spatial zero-inflated Beta models $\mathcal{M}_{2}$ (left) and $\mathcal{M}_{3}$ (right) fit on the {Restionaceae} data in Baviaanskloof.}
    \centering
    \includegraphics[scale = 0.4]{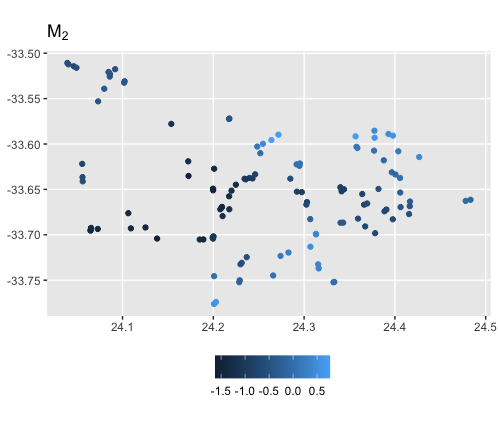} \quad
    \includegraphics[scale = 0.4]{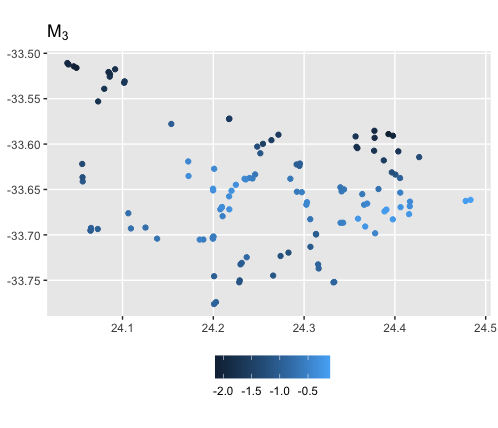}
    \label{fig:restio_res}
\end{figure}

\begin{figure}[h]
    \caption{Plots of functions $\mu$ vs $c_{nu,p}(\mu)$ for various $\nu$ (colors) and $p$ (panels).}
    \centering
    \includegraphics[scale= 0.25]{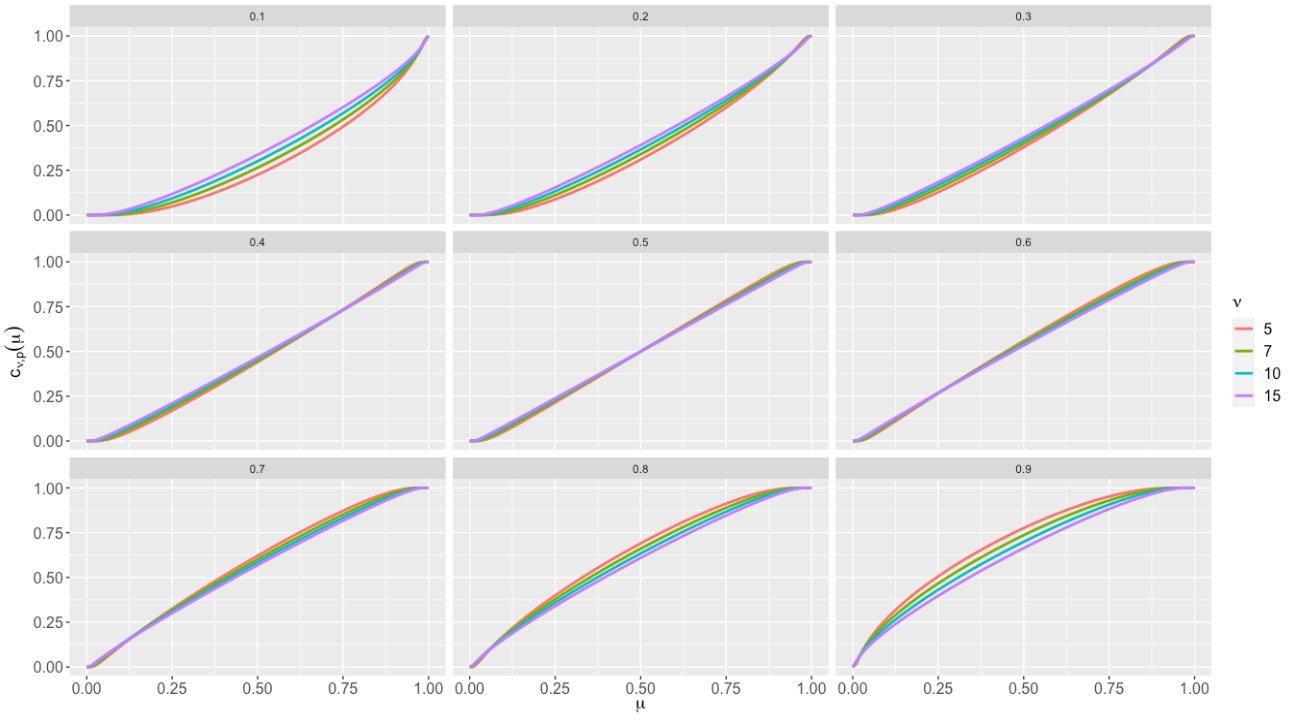}
    \label{fig:plots_mu_c}
\end{figure}

\begin{figure}[h]
    \caption{Plots of $c_{\nu,p}(\mu)$ vs. $\mu$ (in blue, similar to Fig. S\ref{fig:plots_mu_c}), and approximations obtained using the class of curves in Eq. 6 in main text (red), for various $\nu$ (columns) and $p$ (rows). The curves are nearly indistinguishable, implying little sensitivity to the choice of $c$ (equivalently, $a$).}
    \centering
    \includegraphics[scale = 0.35]{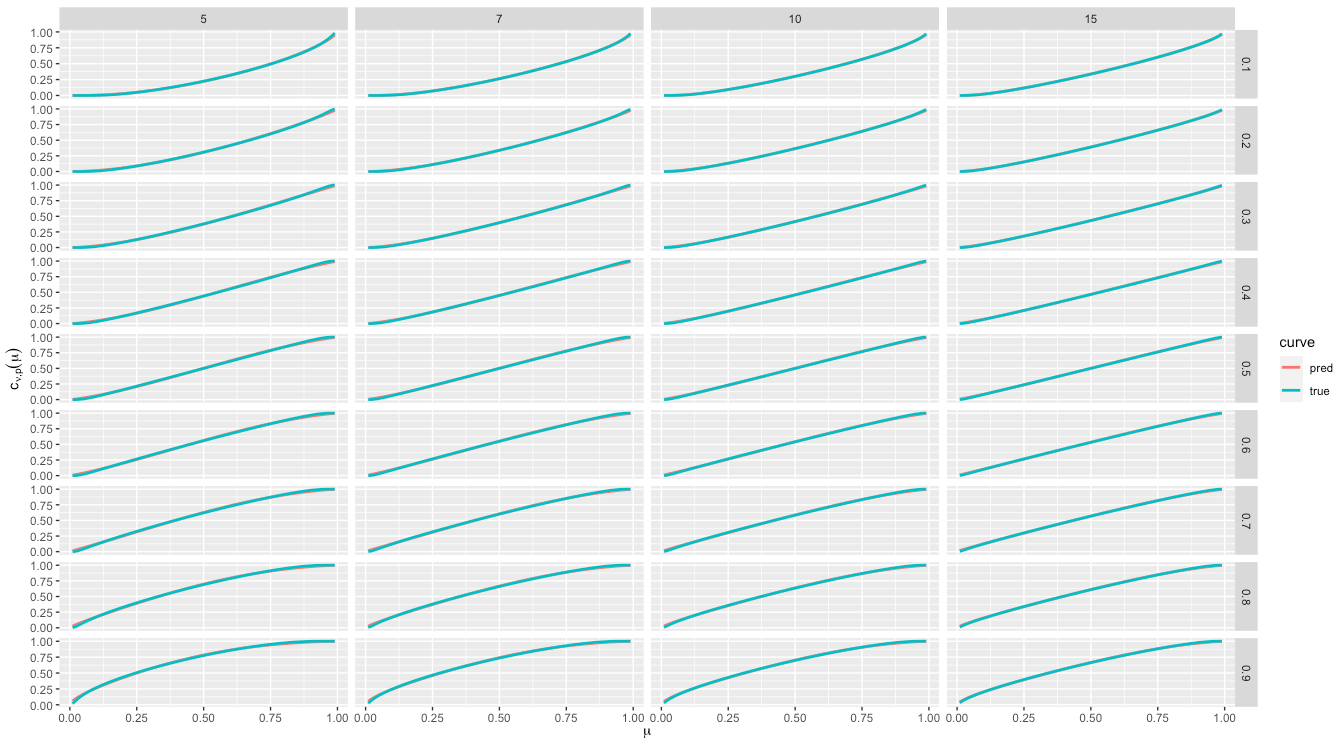}
    \label{fig:curve_fit}
\end{figure}

\clearpage
\section{Supplementary tables}

\begin{table}[ht]
    \caption{Our choices of prior distributions for this paper. Here, $g_{0}$, $\delta_{0}$, and $b$ vary depending on the context (simulation vs application). The notation for the Normal distribution denotes mean and standard deviation. Dashes denote parameters that are not applicable for the given model. Note for all $\mathcal{M}$ models, $a$ and (for the spatial models, $\phi$) are assumed to be held fixed.}
    \centering
    \begin{tabular}{|l|c|c|c|c|c|}
    \hline 
    Parameter & BEZI & LCEB & $\mathcal{M}_{1}$ & $\mathcal{M}_{2}$ & $\mathcal{M}_{3}$ \\
    \hline
        $\gamma_{0}$ & $N(0, 10)$ & - & $N(g_{0}, 0.25)$ & $N(0, 10)$ & $N(g_{0}, 0.25)$ \\
        $\gamma_{j}, j \neq 0$  & $N(0, 10)$ & - & $N(0, 10)$ & $N(0, 10)$ & $N(0, 10)$ \\
        $\delta_{0}$ &  $N(0, 10)$&  $N(0, 10)$& $N(0, 10)$& $N(d_{0}, 0.25)$& $N(0, 10)$\\
        $\delta_{k}, k \neq 0$ & $N(0, 10)$ &  $N(0, 10)$&  $N(0, 10)$& $N(0, 10)$ & $N(0, 10)$ \\
        $\psi$ & $N(0, 10)$ & $N(0, 10)$& $N(0, 10)$ & $N(0, 10)$ & $N(0, 10)$ \\
        $\sigma^{2}$ & -& -& - & $Inv. Ga(0.5, 0.5)$ & $Inv. Ga(0.5, 0.5)$\\
        \hline
    \end{tabular}
    \label{tab:prior_choices}
\end{table}

\begin{table}[ht]
\caption{Posterior mean and 95\% credible intervals for parameters in Simulations 1.1-1.3. }
    \centering
    \resizebox{\textwidth}{!}{
\begin{tabular}{|r|c|ccc|c|ccc|c|ccc|}
  \hline
  &\multicolumn{4}{c|}{Simulation 1.1}&\multicolumn{4}{c|}{Simulation 1.2}&\multicolumn{4}{c|}{Simulation 1.3} \\\hline
    &\multicolumn{4}{c|}{\footnotesize{27\% unsuitable, 27\% ecological}}&\multicolumn{4}{c|}{\footnotesize{12\% unsuitable, 29\% ecological}}&\multicolumn{4}{c|}{\footnotesize{40\% unsuitable, 10\% ecological}} \\\hline
 & true & mean & 2.5\% & 97.5\%& true & mean & 2.5\% & 97.5\% & true & mean & 2.5\% & 97.5\% \\ 
 \hline
$\gamma_0$  & -1.25 & -1.280 & -1.606 & -0.936 & -2.50 & -2.526 & -2.978 & -2.097 & -0.50 & -0.604 & -0.834 & -0.351 \\ 
  $\gamma_1$ & 0.75 & 0.665 & 0.447 & 0.977  & 0.75 & 0.705 & 0.313 & 1.139& 0.75 & 0.716 & 0.534 & 0.935\\ 
  $\delta_0$ & 0.50 & 0.443 & 0.326 & 0.590 & 0.50 & 0.463 & 0.335 & 0.596 & 1.25 & 1.255 & 1.099 & 1.404 \\
  $\delta_1$ &  -0.50 & -0.481 & -0.558 & -0.395 & -0.50 & -0.487 & -0.558 & -0.410& -0.50 & -0.495 & -0.579 & -0.416 \\ 
  $\psi$ & 1.50 & 1.547 & 1.231 & 1.835& 1.50 & 1.579 & 1.374 & 1.776& 1.50 & 1.589 & 1.368 & 1.849 \\
   \hline
\end{tabular}}
\label{tab:m1_post_sums}
\end{table}

\begin{table}[h]
    \caption{Simulation for data generated under the BEZI model. $R^2$ and AUC for classifying between zero and positive, and $CRPS_h$ for predictive distribution of positive observations for various simulations. Simulations vary by amount of zeros in the data, as well the value of $\delta_{0}$.}
    \centering
    \resizebox{\textwidth}{!}{%
    \begin{tabular}{|c|ccc|ccc|ccc|}
    \hline
    &\multicolumn{3}{c|}{$R^2$}&\multicolumn{3}{c|}{AUC} &\multicolumn{3}{c|}{$CRPS_h$} \\ \hline
   \small{(\% 0, $\delta_{0}$)} & BEZI & LCEB & $\mathcal{M}_{1}$  & BEZI & LCEB & $\mathcal{M}_{1}$ & BEZI & LCEB & $\mathcal{M}_{1}$  \\ \hline
    25\%, 1 & 0.023 & 0.002 & \textbf{0.024} & 0.598 & 0.562 & \textbf{0.599} & 0.076 & 0.106 & \textbf{0.053}\\ 
    25\%, -1  & \textbf{0.031} & 0.001 & 0.025 & \textbf{0.622} & 0.547 & 0.575 & \textbf{0.054} & 0.107 & 0.057 \\ 
        \hline \hline
    50\%, 1 & \textbf{0.041} & 0.000 & \textbf{0.041} & \textbf{0.613}& 0.554 & 0.611 & \textbf{0.053} & 0.149 & 0.107 \\ 
    50\%, -1 & \textbf{0.035} & -0.002 & 0.030 & \textbf{0.599} & 0.533 & 0.574 & \textbf{0.054} & 0.107 & 0.073  \\ 
    \hline \hline 
    75\%, 1 & \textbf{0.030} & -0.001 & \textbf{0.030} & \textbf{0.614} & 0.549 & \textbf{0.614} & 0.110 & 0.249 & \textbf{0.054} \\ 
    75\%,  -1& \textbf{0.023} & -0.002 & 0.021 & \textbf{0.590} & 0.563 & 0.578 & \textbf{0.055} & 0.109 & 0.065  \\
    \hline 
    \end{tabular}
    }
    \label{tab:bezi_full_sim}
\end{table}

\begin{table}[h]
    \caption{Simulation for data generated under the censor-only model LCEB. $R^2$ and AUC for classifying between zero and positive, and $CRPS_h$ for predictive distribution of positive observations for various simulations. Simulations vary by amount of zeros in the data.}
    \centering
    \resizebox{\textwidth}{!}{%
    \begin{tabular}{|c|ccc|ccc|ccc|}
    \hline
    &\multicolumn{3}{c|}{$R^2$}&\multicolumn{3}{c|}{AUC} &\multicolumn{3}{c|}{$CRPS_h$} \\ \hline
   \small{\% 0} & BEZI & LCEB & $\mathcal{M}_{1}$  & BEZI & LCEB & $\mathcal{M}_{1}$ & BEZI & LCEB & $\mathcal{M}_{1}$  \\ \hline
    25\% & 0.120 & 0.120 & \textbf{0.122} & \textbf{0.722} & \textbf{0.722} & \textbf{0.722} & 0.085 & 0.083 & \textbf{0.075} \\ 
    50\%  & 0.164 & 0.164 & \textbf{0.165} & \textbf{0.736} &\textbf{0.736} & \textbf{0.736} & 0.140 & 0.091 & \textbf{0.072} \\ 
    75\% & 0.142 & \textbf{0.135} & 0.138  & \textbf{0.735} & \textbf{0.735} & \textbf{0.735} & 0.130 & 0.104 & \textbf{0.085}\\ 
    \hline 
    \end{tabular}
    }
    \label{tab:cens_full_sim}
\end{table}

\begin{table}[ht]
\caption{Posterior summaries for coefficients for simulations where data are generated under $\mathcal{M}_2$, with simulations differing in zero amounts and sources. Simulation 2.1: 20.3\% of data are unsuitable zeros and 28.3\% censored zeros. Simulation 2.2: 30.7\% of data are unsuitable zeros and 11\% censored zeros. Simulation 2.3: 29.5\% of data are unsuitable zeros and 11\% censored zeros.}
\centering
\begin{tabular}{|l|lr|ccc|ccc|}
  \hline
  & & & \multicolumn{3}{c|}{$\mathcal{M}_{2}$}
  & \multicolumn{3}{c|}{$\mathcal{M}_{1}$}\\ \hline
 & & true & mean & 2.5\% & 97.5\% & mean & 2.5\% & 97.5\% \\ 
  \hline
  \multirow{7}{*}{\begin{sideways}Simulation 2.1\end{sideways}} & 
  $\gamma_0$ & -1.000 & -1.275 & -1.683 & -0.851 & -1.121 & -1.618 & -0.670 \\ 
  & $\gamma_1$ &  0.750 & 0.579 & 0.265 & 1.004 & 0.588 & 0.263 & 1.046 \\ 
  & $\delta_0$ & 1.500 & 1.427 & 1.093 & 1.925 & 0.733 & 0.534 & 0.952 \\ 
 & $\delta_1$ &-1.000 & -0.959 & -1.103 & -0.796 & -0.851 & -1.020 & -0.676 \\ 
  & $\psi$ &  2.000 & 1.926 & 1.664 & 2.187 & 0.930 & 0.707 & 1.156 \\ 
 & $\sigma^{2}_{\eta}$ & 1.000 & 1.016 & 0.782 & 1.352 &  -&-  & - \\ 
  & $\phi_{\eta}$ &15.000 & 19.830 & 19.830 & 19.830 &  -&-  &-\\
   \hline
\multirow{7}{*}{\begin{sideways}Simulation 2.2\end{sideways}} &   
$\gamma_{0}$ &  -1.000 & -1.031 & -1.345 & -0.704 & -1.460 & -1.992 & -0.854 \\ 
  &$\gamma_{1}$ & 0.750 & 1.103 & 0.724 & 1.538 & 1.364 & 0.928 & 2.043 \\ 
 & $\delta_{0}$  &1.500 & 1.431 & 1.136 & 1.820 & 1.187 & 0.993 & 1.409 \\ 
  &$\delta_{1}$ &-1.000 & -1.015 & -1.151 & -0.878 & -0.920 & -1.051 & -0.766 \\ 
&$\psi$ & 2.000 & 2.001 & 1.780 & 2.258 & 1.219 & 0.964 & 1.515 \\ 
& $\sigma^{2}_{\eta}$ & 0.500 & 0.496 & 0.283 & 0.799 &  -& - &-\\
  &$\phi_{\eta}$ & 15.000 & 20.071 & 20.071 & 20.071 &  -&  -&-  \\ 
   \hline
   \multirow{7}{*}{\begin{sideways}Simulation 2.3\end{sideways}} &   
  $\gamma_{0}$ &  -2.000 & -1.374 & -2.110 & -0.629 & -0.298 & -0.745 & 0.174 \\ 
  & $\gamma_{1}$ & 0.750 & 0.475 & 0.031 & 1.172 & 0.247 & -0.038 & 0.624 \\ 
  & $\delta_{0}$  & 1.500 & 1.367 & 1.006 & 1.796 & 0.291 & 0.056 & 0.553 \\ 
  & $\delta_{1}$ & -1.000 & -0.995 & -1.151 & -0.831 & -0.821 & -0.980 & -0.637 \\ 
  & $\psi$ & 2.000 & 1.927 & 1.467 & 2.498 & 1.337 & 1.085 & 1.655 \\ 
    & $\sigma^{2}_{\eta}$& 2.000 & 2.010 & 1.451 & 2.704 & - &  -& - \\ 
  & $\phi_{\eta}$& 15.000 & 20.311 & 20.311 & 20.311 &  -&  -&-  \\ 
  \hline
\end{tabular}
\label{tab:re1_post_sum}
\end{table} 

\begin{table}[h]
    \caption{Metrics for comparing models $\mathcal{M}_1$ and $\mathcal{M}_2 $ for data generated under $\mathcal{M}_2$ with $\sigma^2_{\eta} = 1$ and $\phi_{\eta} = 15$ in $(0,50) \times (0,50)$ region. Comparison metrics of area under the ROC curve (AUC) for determining source of zero, Tjur's $R^2$ for classifying between zero and positive, and $CRPS_h$ and $CRPS_f$ for predictive distributions. Simulations vary by amount of zeros in the data, as well as the proportions of zeros arising from each source (\% unsuitable and \% unsuitable zeros), controlled largely by varying intercepts $\gamma_{0}$ and $\delta_{0}$. Larger AUC and $R^2$ and smaller $CRPS$ indicate superior  performance (in bold).}
    \centering
    \begin{tabular}{|c|cc|cc|cc|cc|}
    \hline
    &\multicolumn{2}{c|}{AUC} &\multicolumn{2}{c|}{$R^2$}&\multicolumn{2}{c|}{$CRPS_h$} &\multicolumn{2}{c|}{$CRPS_f$} \\ \hline
   \small{(\% deg. 0, \% Beta 0)}  & $\mathcal{M}_{1}$  & $\mathcal{M}_{2}$ & $\mathcal{M}_{1}$  & $\mathcal{M}_{2}$
    & $\mathcal{M}_{1}$  & $\mathcal{M}_{2}$  & $\mathcal{M}_{1}$  & $\mathcal{M}_{2}$\\ \hline
    (14\%, 9\%) & 0.815 & \textbf{0.877} & 0.116 & \textbf{0.173} & 0.064 & \textbf{0.053} & 0.233& \textbf{0.228} \\ 
    (9\%, 10\%) & 0.835 & \textbf{0.893} & 0.116 & \textbf{0.192} & 0.064 & \textbf{0.053} & 0.213 & \textbf{0.207} \\ 
    (6\%, 13\%)& 0.807 &  \textbf{0.878} & 0.118 &\textbf{0.201} &0.060 &\textbf{0.051} &0.188 & \textbf{0.181} \\ 
    \hline      \hline
    (33\%, 16\%)& 0.765 & \textbf{0.822} & 0.106 & \textbf{0.156} & 0.078 & \textbf{0.065} & 0.317 & \textbf{0.316} \\ 
    (21\%, 23\%)& 0.786 & \textbf{0.830} & 0.151 & \textbf{0.236} & 0.083 & \textbf{0.067} & 0.347 & \textbf{0.345} \\ 
    (15\%, 33\%)& 0.719& \textbf{0.768} & 0.188 & \textbf{0.294} & 0.093& \textbf{0.075} & 0.462 & \textbf{0.460} \\ 
    \hline \hline
    (66\%, 9\%)& 0.767& \textbf{0.804} & 0.068 & \textbf{0.084} & 0.181 & \textbf{0.081} & \textbf{0.224} & 0.227 \\ 
    (40\%, 33\%) & 0.721 & \textbf{0.754} & 0.126 & \textbf{0.174} & 0.093 & \textbf{0.079} & \textbf{0.364} & 0.370 \\ 
    (16\%, 60\%) & 0.632& \textbf{0.643} & 0.200 & \textbf{0.272} & 0.111 & \textbf{0.096} & \textbf{0.851} & 0.870 \\ 
    \hline
    \end{tabular}
    \label{tab:sims_m2}
\end{table}


\begin{table}[ht]
\caption{Posterior summaries for coefficients for simulations where data are generated under $\mathcal{M}_3$, with differing in zero amounts and sources. Simulation 3.1: 34\% of data are unsuitable zeros and 11.3\% censored zeros. Simulation 3.2: 12\% of data are unsuitable zeros and 25.3\% censored zeros. Simulation 3.3: 22.3\% of data are unsuitable zeros and 26.7\% censored zeros.}
\centering
\begin{tabular}{|l|lr|ccc|ccc|}
  \hline
  & & & \multicolumn{3}{c|}{$\mathcal{M}_{3}$}
  & \multicolumn{3}{c|}{$\mathcal{M}_{1}$}\\ \hline
 & & true & mean & 2.5\% & 97.5\% & mean & 2.5\% & 97.5\% \\ 
  \hline
  \multirow{7}{*}{\begin{sideways}Simulation 3.1\end{sideways}} & 
  $\gamma_{0}$ &  -1.000 & -0.972 & -1.347 & -0.507 & -0.863 & -1.108 & -0.592 \\ 
  & $\gamma_{1}$ &   0.750 & 0.736 & 0.427 & 1.147 & 0.715 & 0.458 & 1.077 \\ 
  & $\delta_{0}$ &  1.500 & 1.414 & 1.246 & 1.593 & 1.351 & 1.185 & 1.523 \\ 
  & $\delta_{1}$ &  -1.000 & -0.987 & -1.085 & -0.877 & -0.984 & -1.077 & -0.866 \\ 
  & $\psi$ & 2.000 & 2.075 & 1.837 & 2.324 & 1.978 & 1.714 & 2.303 \\ 
    & $\sigma^{2}_{\tau}$&  1.000 & 0.930 & 0.422 & 1.520 &-  & - & - \\ 
  & $\phi_{\tau}$& 15.000 & 20.311 & 20.311 & 20.311 &  -& - &-  \\ 
   \hline
\multirow{7}{*}{\begin{sideways}Simulation 3.2\end{sideways}} &   
$\gamma_{0}$ & -2.000 & -2.006 & -2.391 & -1.574 & -1.909 & -2.197 & -1.559 \\ 
  &$\gamma_{1}$ &0.750 & 0.255 & -0.216 & 0.838 & 0.178 & -0.256 & 0.742 \\ 
 & $\delta_{0}$  & 1.000 & 1.019 & 0.874 & 1.176 & 1.009 & 0.871 & 1.156 \\ 
  &$\delta_{1}$ & -1.000 & -0.962 & -1.045 & -0.873 & -0.969 & -1.050 & -0.865 \\ 
&$\psi$ & 2.000 & 2.023 & 1.793 & 2.291 & 2.020 & 1.797 & 2.266 \\ 
  &$\sigma^{2}_{\tau}$ &  0.500 & 0.497 & 0.141 & 1.172 & - &-  &-  \\ 
    &$\phi_{\tau}$ & 15.000 & 20.311 & 20.311 & 20.311 & - & - &  -\\ 
   \hline
   \multirow{7}{*}{\begin{sideways}Simulation 3.3\end{sideways}} &   
  $\gamma_{0}$ & -2.000 & -1.949 & -2.366 & -1.460 & -1.661 & -1.960 & -1.280 \\ 
  & $\gamma_{1}$  & 0.750 & 0.911 & 0.483 & 1.452 & 0.653 & 0.218 & 1.140 \\ 
  & $\delta_{0}$ & 1.000 & 0.895 & 0.788 & 1.017 & 0.819 & 0.692 & 0.987 \\ 
  & $\delta_{1}$ & -1.000 & -1.090 & -1.189 & -0.973 & -1.080 & -1.165 & -0.989 \\ 
  & $\psi$ & 2.000 & 2.164 & 1.969 & 2.402 & 2.058 & 1.838 & 2.354 \\ 
    & $\sigma^{2}_{\tau}$&  2.000 & 4.215 & 1.512 & 6.643 & - & - & - \\ 
  & $\phi_{\tau}$& 15.000 & 20.311 & 20.311 & 20.311 &  -& - & - \\ 
  \hline
\end{tabular}
\label{tab:re2_post_sum}
\end{table} 

\begin{table}[h]
    \caption{Metrics for comparing models $\mathcal{M}_1$ and $\mathcal{M}_3$ for data generated under $\mathcal{M}_3$, with $\sigma^{2}_{\tau} = 1$ and $\phi_{\tau} = 15$ in a $(0, 50) \times (0,50)$ region. Comparison metrics of area under the ROC curve (AUC) for determining source of zero, Tjur's $R^2$ for classifying between zero and positive, and $CRPS_h$ and $CRPS_f$ for predictive distributions. Simulations vary by amount of zeros in the data as well as the proportions of zeros arising from each source (\% unsuitable and \% unsuitable zeros), largely controlled by varying the intercepts $\gamma_{0}$ and $\delta_{0}$. Larger AUC and $R^2$ and smaller $CRPS$ indicate superior  performance (in bold).}
    \centering
    \begin{tabular}{|c|cc|cc|cc|cc|}
    \hline
    &\multicolumn{2}{c|}{AUC} &\multicolumn{2}{c|}{$R^2$}&\multicolumn{2}{c|}{$CRPS_h$} &\multicolumn{2}{c|}{$CRPS_f$} \\ \hline
   \small{(\% deg. 0, \% Beta 0)}  & $\mathcal{M}_{1}$  & $\mathcal{M}_{3}$ & $\mathcal{M}_{1}$  & $\mathcal{M}_{3}$
    & $\mathcal{M}_{1}$  & $\mathcal{M}_{3}$  & $\mathcal{M}_{1}$  & $\mathcal{M}_{3}$\\ \hline
    (20\%, 8\%) & 0.908 & \textbf{0.914} & 0.176 & \textbf{0.202} & 0.049 & \textbf{0.047} & 0.242 & \textbf{0.241} \\ 
     (11\%, 10\%) & 0.916 & \textbf{0.919} & 0.233 & \textbf{0.244} & \textbf{0.048} & \textbf{0.048} & \textbf{0.216} & \textbf{0.216} \\ 
     (6\%, 13\%) & 0.915 &  \textbf{0.916} & 0.278 &\textbf{0.283} & \textbf{0.049} &\textbf{0.049} & \textbf{0.197} & \textbf{0.197} \\ 
    \hline      \hline
    (36\%, 14\%) & 0.857 & \textbf{0.871} & 0.181 & \textbf{0.222} & 0.058 & \textbf{0.056} & 0.300 & \textbf{0.295} \\ 
    (23\%, 24\%) & 0.800 & \textbf{0.821} & 0.316 & \textbf{0.327} & 0.065 & \textbf{0.062} & 0.380 & \textbf{0.379} \\ 
    (20\%, 33\%) & 0.792& \textbf{0.813} & 0.350 & \textbf{0.360} & 0.065 & \textbf{0.064} & 0.432 & \textbf{0.422} \\ 
    \hline \hline
    (65\%, 9\%) & 0.838& \textbf{0.854} & 0.093 & \textbf{0.136} & 0.061 & \textbf{0.059} & 0.243 & \textbf{0.233} \\ 
    (49\%, 25\%) & 0.772 & \textbf{0.792} & 0.249 & \textbf{0.269} & 0.067 & \textbf{0.066} & 0.382 & \textbf{0.380} \\ 
    (21\%, 58\%) & 0.658& \textbf{0.673} & 0.400 & \textbf{0.410} & 0.073 & \textbf{0.071} & 0.690 & \textbf{0.685}\\ 
    \hline
    \end{tabular}
    \label{tab:sims_m3}
\end{table}

\begin{table}[ht]
\caption{Posterior summaries of parameters for the two spatial models fit to the Restionaceae percent cover data in the Baviaanskloof region. Note that rainfall concentration and minimum annual July temperature were not found to be significant. $\mathcal{M}_2$ includes the spatial effect in $\mu_i$, with the prior for $\delta_0$ centered at 0.5. $\mathcal{M}_3$ includes the effect in $\pi_i$, with the prior for $\gamma_{0}$  centered at -0.5.}
\centering
\begin{tabular}{|l|ccc|ccc|}
  \hline
  & \multicolumn{3}{c|}{$\mathcal{M}_2$} & \multicolumn{3}{c|}{$\mathcal{M}_3$} \\ \hline
 & mean & 2.5\% & 97.5\%  & mean & 2.5\% & 97.5\% \\ 
  \hline
$\gamma_0$: Intercept &  -2.117 & -2.777 & -1.338 & -0.911 & -1.303 & -0.473 \\ 
 $\gamma_1$: mean\_ann\_pan & - & - &- & 1.167 & 0.054 & 2.367 \\ \hline
  $\delta_{0}$: Intercept & 1.029 & 0.762 & 1.465 & 0.543 & 0.201 & 1.045 \\ 
  $\delta_{1}$:  mean\_ann\_precip  & -1.474 & -1.859 & -1.088 & -0.974 & -1.305 & -0.589 \\ 
 $\psi$ & 1.238 & 0.867 & 1.707 & 0.662 & 0.320 & 1.095 \\ \hline
  $\sigma^{2}$ & 0.967 & 0.582 & 1.629 & 1.167 &0.732 & 2.143 \\ 
   \hline
\end{tabular}
\label{tab:restio_spatial_postsum}
\end{table}

\begin{table}[h]
    \caption{Model comparison metrics for the five models (BEZI, LCEB, $\mathcal{M}_{1}$, $\mathcal{M}_{2}$,  $\mathcal{M}_{3}$) fit to Crassulaceae data in Baviaanskloof region. The $\mathcal{M}$ models are fit with $c$ fixed at different values. Best performer for each metric in bold.}
    \centering
    \begin{tabular}{|l|cccc|}
    \hline
    & $R^2$ & AUC & $CRPS_h$ & $CRPS_f$ \\ \hline
    BEZI & 0.126 & 0.709 & 0.178 & - \\
    LCEB $(c = 0.1)$ & 0.159 & 0.719 & 0.162 & - \\
    LCEB $(c = 0.3)$ & 0.174 & 0.718 & 0.123 & - \\
    LCEB $(c = 0.5)$ & 0.180 & 0.717 & 0.182 & -\\
    $\mathcal{M}_1$ $(c = 0.1)$ & 0.165 & 0.726 & 0.160 & 0.350  \\
    $\mathcal{M}_1$ $(c = 0.3)$ & 0.170 & 0.716 & 0.124 & 0.330  \\
    $\mathcal{M}_1$ $(c = 0.5)$ & 0.172 & 0.720 & 0.089 & 0.299  \\
   $\mathcal{M}_2$ $(c=0.1)$ & 0.218 & 0.760 & 0.143 & 0.452 \\
      $\mathcal{M}_2$ $(c=0.3)$ & 0.219 & 0.750 &  0.111 & 0.308 \\
   $\mathcal{M}_2$ $(c=0.5)$ & \textbf{0.225} & \textbf{0.766} & \textbf{0.080} & \textbf{0.277} \\
   $\mathcal{M}_3$ $(c = 0.1)$ & 0.154 & 0.716 & 0.159 & 0.378 \\
   $\mathcal{M}_3$ $(c = 0.3)$ & 0.166 & 0.712 & 0.124 & 0.372 \\
   $\mathcal{M}_3$ $(c = 0.5)$ & 0.156 & 0.714 & 0.089 & 0.358 \\
   \hline
    \end{tabular}
    \label{tab:crass_mods}
\end{table}

\begin{table}[h]
    \caption{Model comparison metrics for the five models (BEZI,  LCEB, $\mathcal{M}_{1}$, $\mathcal{M}_{2}$, $\mathcal{M}_{3}$) fit to Restionaceae data in Baviaanskloof region. The $\mathcal{M}$ models are fit with $c$ fixed at different values. Best performer for each metric in bold.}
    \centering
    \begin{tabular}{|l|cccc|}
    \hline
    & $R^2$ & AUC & $CRPS_h$ & $CRPS_f$ \\ \hline
    BEZI & 0.155 & 0.728 & 0.163 & - \\
    LCEB $(c = 0.1)$ & 0.341 & 0.833 & 0.192 & - \\
    LCEB $(c = 0.3)$ & 0.351 & 0.884 & 0.148 & - \\
    LCEB $(c = 0.5)$ &  0.354 & 0.885 & 0.105 & -\\
    $\mathcal{M}_1$ $(c = 0.1)$ & 0.289 & 0.867 & 0.193 & 0.404  \\
    $\mathcal{M}_1$ $(c = 0.3)$ & 0.288 & 0.867 & 0.185 & 0.391  \\
    $\mathcal{M}_1$ $(c = 0.5)$ & 0.278 & 0.860 & 0.183 & 0.419  \\
   $\mathcal{M}_2$ $(c=0.1)$ & 0.367 & 0.\textbf{908} & 0.148 & 0.415 \\
$\mathcal{M}_2$ $(c=0.3)$ & 0.416 & \textbf{0.908} &  0.119 & 0.443 \\
   $\mathcal{M}_2$ $(c=0.5)$ & \textbf{0.432} & \textbf{0.908} & \textbf{0.086} & 0.455\\
   $\mathcal{M}_3$ $(c = 0.1)$ & 0.278 & 0.871 & 0.161 & \textbf{0.303} \\
   $\mathcal{M}_3$ $(c = 0.3)$ & 0.300 & 0.870 & 0.127 & 0.320 \\
   $\mathcal{M}_3$ $(c = 0.5)$ & 0.310 & 0.866 & 0.091 & 0.327 \\
   \hline
    \end{tabular}
    \label{tab:rest_mods}
\end{table}

\end{document}